\documentclass[11pt]{article}

\usepackage{a4wide}
\usepackage[fleqn]{amsmath}
\usepackage{hyperref}
\usepackage{fourier} 
\usepackage[scaled=0.875]{helvet} 

\RequirePackage{amsmath,amssymb,amsxtra,amsthm}

\RequirePackage{mathrsfs}

\RequirePackage{setspace}
\RequirePackage{array}
\RequirePackage{booktabs}

\RequirePackage{braket}

\RequirePackage[pdftex,final]{graphicx}
\usepackage{graphicx}
\usepackage{subcaption}
\usepackage{float}

\RequirePackage{colortbl}
\RequirePackage{xcolor}
\colorlet{titlerowcolor}{gray!15}

\usepackage{cite}

\numberwithin{equation}{section}
\numberwithin{table}{section}
\numberwithin{figure}{section}

\author{
  \begin{minipage}{1.00\linewidth}
    \vspace{1cm}
    \begin{center}
      \begin{small}
        \textbf{Carlo Angelantonj$^1$, Ioannis Florakis$^2$ and Giorgio Leone$^1$}
        \\ \vspace{1cm}
        ${}^1$ {\em Dipartimento di Fisica, Universit\`a di Torino and INFN Sezione di Torino}
        \\
        {\em Via Pietro Giuria 1, I-10125 Torino}
        \\
        ${}^2$ {\em Department of Physics, University of Ioannina, GR-45110, Ioannina}
     \end{small}
    \end{center}
    \vspace{1cm}
  \end{minipage}
}

\date{}

\title{\vspace{3cm}
  \begin{huge} \textbf{Tachyons and Misaligned Supersymmetry in Closed String Vacua} 	
  \end{huge}
  \\ \vspace{.7cm}
}

\begin{document}

\begin{titlepage}
  \maketitle
  \thispagestyle{empty}

  \vspace{-14cm}
  \begin{flushright}
   \end{flushright}

  \vspace{11cm}

  \begin{center}
    \textsc{Abstract}\\
  \end{center}

In a remarkable paper, Dienes discovered that the absence of physical tachyons in closed string theory is intimately related to oscillations in the net number of bosonic minus fermionic degrees of freedom, a pattern predicted by an underlying misaligned supersymmetry. The average of these oscillations was linked to an exponential growth controlled by an effective central charge $C_\text{eff}$ smal\-ler than the expected inverse Hagedorn temperature. Dienes also conjectured that $C_\text{eff}$ should vanish when tachyons are absent.

In this paper, we revisit this problem and show that boson-fermion oscillations are realised even when tachyons are present in the physical spectrum. In fact, we prove that the average growth rate $C_\text{eff}$ is set by the mass of the ``lightest'' state, be it massless or tachyonic, and coincides with the effective inverse Hagedorn temperature of the associated thermal theory. We also provide a general proof that the necessary and sufficient condition for classical stability is the vanishing of the sector averaged sum which implies $C_\text{eff} =0$, in agreement with Dienes' conjecture.

\vfill

{\small
\begin{itemize}
\item[E-mail:] {\tt carlo.angelantonj@unito.it}
\\
{\tt iflorakis@uoi.gr}
\\
{\tt giorgio.leone@unito.it}

\end{itemize}
}

\end{titlepage}

\setstretch{1.1}


\tableofcontents

\section{Introduction}

Superstring vacua are typically unstable when space-time supersymmetry is absent. The instability can be ascribed to the emergence of tadpoles for scalar fields or, even worse, to the presence of tachyons in the classical spectrum. In the first case, much progress has been made in the last years in the determination of the corrected geometry which can describe, for instance, spontaneous compactification or cosmological evolution \cite{Dudas:2000ff, DeWolfe:2001nz, Gubser:2001zr, Dudas:2002dg, Dudas:2010gi, Mourad:2016xbk, Basile:2018irz, Mourad:2021qwf, Mourad:2021roa, Mourad:2022loy, Antonelli:2019nar, Basile:2020mpt, Basile:2021krk, Raucci:2022jgw, Raucci:2022bjw,Baykara:2022cwj}. The second case, instead, is much less under control. In fact, although the condensation of open-string tachyons is  well understood \cite{Sen:1999nx,Schnabl:2005gv}, very little is known about the rolling of closed-string ones. Moreover, in critical strings tadpoles typically emerge at higher genus and, as such, induce quantum corrections to the classical moduli space. In this case, perturbation theory is relatively under control at least up to the order associated to the tadpole itself. On the contrary, tree-level tachyons imply that the string vacuum cannot be trusted even classically, and it is impossible to extract any meaningful information from it.  Therefore, if one insists that a non-supersymmetric vacuum ought to be predictive, at least classically, one is bound to consider only string configurations where tachyonic excitations are absent. This turns out to be rather non-trivial to achieve and requires a precise correlation in the growth rates of massive bosonic and fermionic string states, since these two features are strictly related by modular invariance. 

The interplay between massive states and tachyons is an inherent property of conformal field theory. In particular, in \cite{Kutasov:1990sv}  it was shown that the infra-red finiteness of the one-loop vacuum energy of closed oriented strings implies the overall cancellation 
\begin{equation}
\lim_{\Lambda \to \infty} \sum_{n} \left( d_\text{B} (n ) - d_\text{F} (n )\right) \, e^{-4\pi n /\Lambda^2} \equiv  \lim_{\Lambda \to \infty} \sum_{n} d  (n )  \, e^{-4\pi n /\Lambda^2} 
= 0\,, \label{KutSei}
\end{equation}
between the total bosonic $d_\text{B} (n )$ and fermionic $d_\text{F} ( n)$ degrees of freedom, at mass-squared $n$. This property goes under the name {\em asymptotic supersymmetry}, since it is reminiscent of the Bose-Fermi degeneracy one typically encounters when supersymmetry is present. We should stress, however, that string theory differs quite drastically from field theory, since one cannot identify a finite energy scale above which supersymmetry is restored.

A more refined analysis based on the Rankin-Selberg-Zagier transform \cite{Rank, Selb, Zagier2} done in \cite{Angelantonj:2010ic}, actually showed that, when physical tachyons are absent and thus the one-loop vacuum energy $\Omega$ is finite, the distribution of bosonic and fermionic degrees of freedom follows a precise pattern dictated by
\begin{equation}
\sum_{n}  d (n) \, e^{-4\pi n /\Lambda^2} \simeq \Lambda^{2-D}\, \frac{3\, \Omega}{\pi} + \sum_{\zeta^* (\rho ) =0} C_\rho \, \Lambda^{\rho_1-D}\, \cos \left(\rho_2 \, \log\Lambda + \varphi_\rho \right) \,, \label{ACER}
\end{equation}
where the sum in the RHS runs over the zeroes $\rho =\rho_1 + i \rho_2$ of the dressed Riemann zeta-function, $\zeta^* (s) = \pi^{-s/2}\, \Gamma (s/2) \, \zeta (s)$, and $C_\rho$ and $\varphi_\rho$ are model-dependent real constants. Indeed, this expression implies that not only must the full string spectrum in $D>2$ non-compact space-time dimensions enjoy {\em asymptotic supersymmetry} but, for finite large $\Lambda$ the excess of bosonic and fermionic degrees of freedom must alternate with a frequency dictated by the non-trivial zeroes of the Riemann zeta-function. This oscillatory behaviour of a classically stable string spectrum was actually discovered already in \cite{Dienes:1994np}, where it was named {\em misaligned supersymmetry}. Notice that the analysis of \cite{Kutasov:1990sv} and \cite{Angelantonj:2010ic} crucially rely on the finiteness of the one-loop modular integral, and thus  the results \eqref{KutSei} and \eqref{ACER} are {\em necessary conditions} for the tree-level stability of a string vacuum, {\em i.e.} for the absence of physical tachyons. Whether these conditions are also sufficient is an open problem which cannot be addressed using the techniques in \cite{Kutasov:1990sv,Angelantonj:2010ic}. 

\begin{figure}[h!]
	\centering
	\includegraphics[width=9cm]{"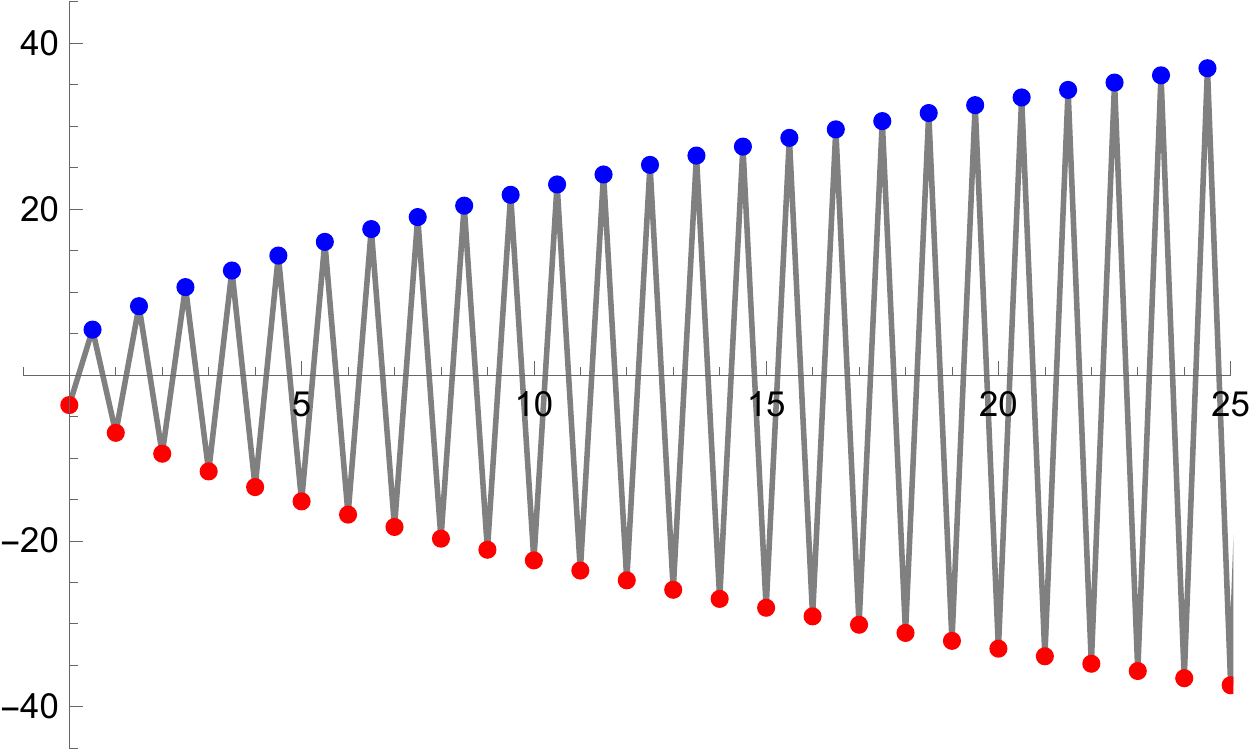"}
	\caption{The {\em signed} logarithm of the net number of degrees of freedom at each mass level for the non-tachyonic $\text{SO}(16) \times \text{SO}(16)$ heterotic string in $D=10$ highlighting the presence of misaligned supersymmetry. Positive (negative) contributions are ascribed to the excess of bosonic (fermionic) states.}
	\label{KDfig1}
\end{figure}

A somehow different route to relate the distribution of degrees of freedom to the classical stability of closed-string vacua was followed by Dienes in the remarkable paper \cite{Dienes:1994np}. There, he observed that the degrees of freedom of non-tachyonic, non-supersymmetric vacua follow a precise pattern as in Figure \ref{KDfig1}, where the net excess between bosons and fermions alternates as the mass-level increases\footnote{Clearly, in the case of unbroken supersymmetry there is an exact degeneracy between bosonic and fermionic states and, therefore, no oscillations are observed in the full spectrum.}. He then observed that the presence of oscillations can be traced back to the growth property of a special quantity, termed the {\em sector averaged sum}, to be introduced shortly. 

In a given string background, the partition function admits the double power-series expansion
\begin{equation}
{\mathscr Z} = \sum_{n,m} \sum_{a,b} {\mathscr N}_{ab}\, d_{ab} (n,m)\, q^{n} \bar q^m \,,
\end{equation}
with $d_{ab} (n,m)$ counting the net number of bosonic minus fermionic states with left mass-squared $n$ and right mass-squared $m$ in the sector $ab$. The matrix ${\mathscr N}_{ab}$ enforces the GSO projection and builds the string vacuum. Propagating degrees of freedom in the $ab$ sector are selected by the level matching condition which identifies the left-moving and right-moving masses, and are then counted by the integral functions $d_{ab} (n) \equiv d_{ab} (n,n )$, whose domain is a countable set associated to the string oscillators, Kaluza-Klein momenta and/or windings. Two-dimensional conformal invariance and modularity imply a universal exponential growth of $d_{ab} (n)$, 
\begin{equation}
d_{ab} (n) \sim e^{C_\text{tot} \sqrt{n}}\,, \label{univgrH}
\end{equation}
controlled by the {\em total central charge} $C_\text{tot} = 4 \pi (\sqrt{c_\text{L}/24} + \sqrt{c_\text{R}/24} )$, which also defines the inverse Hagedorn temperature\footnote{Of course, the total central charge of the full CFT is $c_\text{L} + c_\text{R}$. However, by abuse of language, we shall refer to $C_\text{tot}$ as the {\em total central charge}, as in \cite{Dienes:1994np}.}. This growth is then inherited by the  total net number of physical states $d (n) =\sum_{a,b} {\mathscr N}_{ab} \, d_{ab} (n) $, obtained by summing over all sectors.

The connection between the asymptotic growth of states and the absence of tachyons cannot be formulated in terms of the physical $d (n)$, but requires the introduction of the continuous enveloping functions $\Phi_{ab} (n)$, which reproduce the $d_{ab} (n)$ for the special values of $n$ associated to the actual masses \cite{Dienes:1994np}. Using these functions, Dienes introduced the {\em sector averaged sum}
\begin{equation}
\langle d (n) \rangle  =\sum_{a,b} {\mathscr N}_{ab}\, \Phi_{ab} (n)\,,
\end{equation}
which discriminates between tachyonic and non-tachyonic vacua. In fact, $\langle d (n) \rangle$ is no-longer bound to exhibit the universal exponential growth \eqref{univgrH}, controlled by the total central charge, but Dienes showed \cite{Dienes:1994np} that in the absence of on-shell tachyons
\begin{equation}
\langle d (n) \rangle \sim e^{C_\text{eff}\, \sqrt{n}}\,,
\end{equation}
with $C_\text{eff} < C_\text{tot}$. The cancellation underlying this slower growth clearly implies the oscillatory behaviour of fig. \ref{KDfig1} observed in the string spectrum.  In \cite{Dienes:1994np} it was also conjectured that the absence of physical tachyons actually implies $C_\text{eff} =0$, a conjecture which was later shown to be true \cite{Cribiori:2020sct} in the case of the ten-dimensional $\text{SO}(16) \times \text{SO}(16)$ heterotic string.  

\begin{figure}[h!]
	\centering
	\includegraphics[width=9cm]{"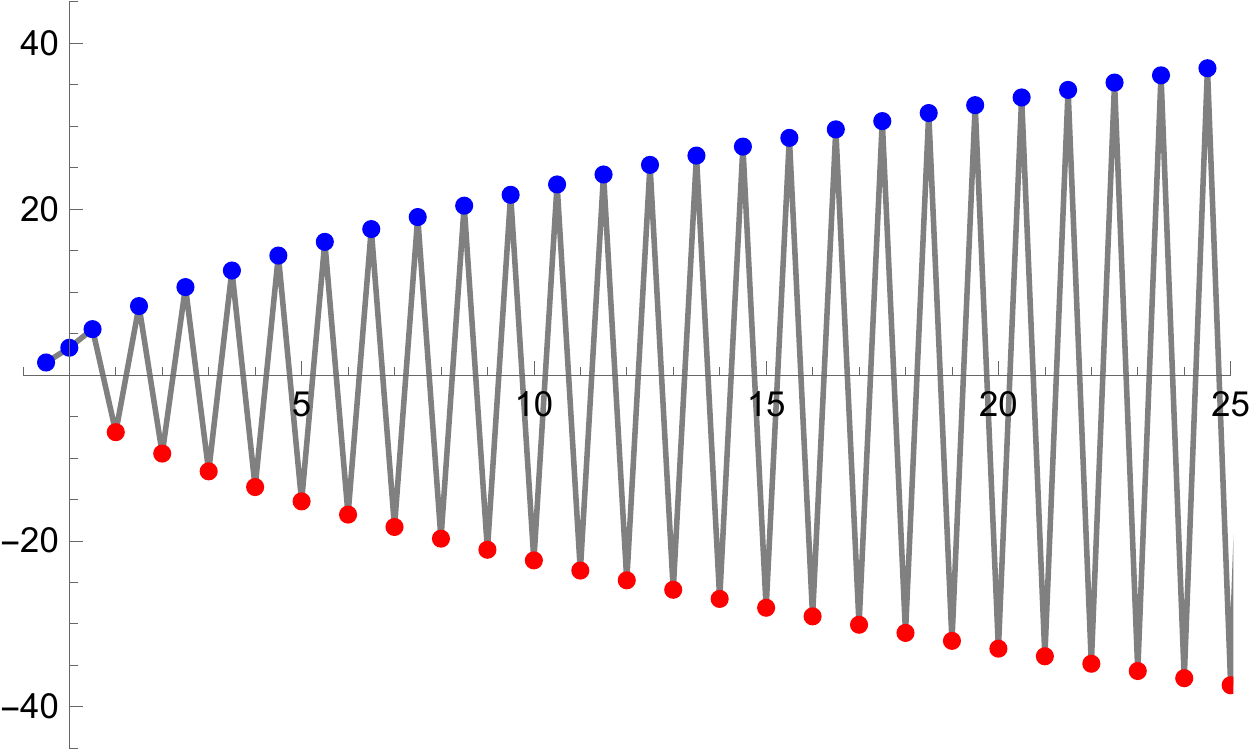"}
	\caption{The {\em signed} logarithm of the net number of degrees of freedom at each mass level for the tachyonic $\text{SO}(16) \times \text{E}_8$ heterotic string in $D=10$. The plots exhibits misaligned supersymmetry even in the presence of physical tachyons. Positive (negative) contributions are ascribed to the excess of bosonic (fermionic) states.}
	\label{figO16E8}
\end{figure}

Simple numerical experiments on non-supersymmetric string vacua, actually reveal a {\em slightly} more complicated story. As stated in \cite{Dienes:1994np}, misaligned supersymmetry is a property emerging in modular-invariant theories that are free of physical tachyons\footnote{We thank K. Dienes for clarifying the original meaning of misaligned supersymmetry.}.  Such theories are of course  IR finite. It predicts a growth rate $C_\text{eff} < C_\text{tot}$ \cite{Dienes:1994np}, a result which is realised through boson/fermion oscillations throughout the string spectrum. Some readers have misinterpreted these results as implying that the presence of boson/fermion oscillations, in and of themselves, is sufficient for the IR finiteness of string theory.   However, this is not the case:  not all boson/fermion oscillations correspond to an underlying misaligned supersymmetry, and thus no statement about IR finiteness can be drawn by looking at the presence of oscillations alone. Indeed, a dramatic example is the $\text{E}_8\times \text{SO}(16)$ non-supersymmetric heterotic theory, where the signed multiplicities follow the oscillatory pattern\footnote{For some early results revealing oscillatory patterns in the presence of physical tachyons see \cite{Sara,Faraggi:2020fwg}.} shown in Figure \ref{figO16E8}. This is so despite the fact that the low-lying spectrum contains a physical tachyon of mass-squared $m^2_T= - 2/\alpha'$ in the representation $(\boldsymbol{16},\boldsymbol{1})$. Also in this case the exponential growth of the sector averaged sum is controlled by a $C_\text{eff} < C_\text{tot}$, thus highlighting a misconception that has propagated in the literature. In fact, $C_\text{tot}$ only sets the growth rate  in purely bosonic theories , like the bosonic string and the type 0A/0B superstrings while, whenever fermions are present in the spectrum, the growth rate is slower than $C_\text{tot}$ \cite{Dienes:1994np}, even when physical tachyons are present, as we show in this work.

In this paper we prove, independently of the details of the string construction, that the growth rate of $\langle d (n) \rangle$ is set by the mass of the {\em lightest} states, whether tachyonic or massless\footnote{By {\em lightest} tachyon we mean the {\em deepest} tachyon, {\em i.e.} the state with largest $|m_\text{tachyon}^2|$.},
\begin{equation}
C_\text{eff} = 4 \pi \sqrt{|\alpha' m^2_\text{lightest} |} \le C_\text{tot}\,, \label{deepestCeff}
\end{equation}
and that the {\em necessary} and {\em sufficient} condition for classical stability, and thus for the presence of misaligned supersymmetry, is the vanishing of the sector averaged sum $\langle d (n) \rangle =0$, also implying the vanishing of the effective central charge $C_\text{eff} = 0$. This proves the conjecture of Dienes for any critical closed-string vacuum, regardless of its CFT realisation. Notice that eq. \eqref{deepestCeff} exhibits the same dependence of the effective Hagedorn temperature on the mass of the deepest tachyon, as discussed in \cite{Dienes:2012dc}, within a corresponding thermal setting.

 The paper is organised as follows. In Section \ref{KV}, we review the exact expression for the degeneracies of states in Rational CFT's (RCFT's). Section \ref{main} deals with the analysis of the sector averaged sum and contains the main result of our work. In Sections \ref{10d} and \ref{ScSc}, we illustrate our results by studying concrete examples in ten and lower dimensions, and we conjecture the presence of first-order phase transitions on the world-sheet of an {\em averaged} CFT associated to the Scherk-Schwarz reduction. 
Our conclusions and perspectives are gathered in Section \ref{concl}. Finally,  in Appendix \ref{appRadexp} we review the computation of the exact degeneracies of vector-valued modular forms, while  Appendix \ref{appScScrational} contains the relevant character decomposition of shift orbifolds of one-dimensional Narain lattices at rational points.

\section{Exact Mass Degeneracies} \label{KV} 

As anticipated, in order to uncover  the origin of misaligned supersymmetry and of the absence of physical tachyons, one needs to study the large-mass behaviour of the string spectrum. For simplicity, we assume that the world-sheet fermions and the compact bosons contribute with a finite number $M$ of characters, both in the holomorphic and anti-holomorphic sectors. In principle, these sets of characters need not  be the same, as for instance, in the heterotic string. Still, for convenience, we shall use the same symbol  $\chi$ to label the contributions $\chi_a$ from the holomorphic sector and $\bar\chi_a$ from the anti-holomorphic one. The $D$ non-compact bosons contribute instead with the Dedekind $\eta$-function, so that the one-loop vacuum energy reads
\begin{equation}
\int_{\mathscr F} d\mu\, {\mathscr Z}  = \int_{\mathscr F} d\mu\, \frac{1}{(\sqrt{\tau_2}\, \bar\eta\, \eta)^{D-2}} \sum_{a,b} \bar \chi_a {\mathscr N}_{ab}\chi_b 
\,, \label{partition}
\end{equation}
where $\tau=\tau_1 + i \tau_2$ is the complex structure modulus of the world-sheet torus, ${\mathscr F}$ is the fundamental domain of $\text{SL} (2;\mathbb{Z})$ and $d\mu = \tau_2^{-2} \, d\tau_1 \, d\tau_2$ is the modular invariant measure.
Again, ${\mathscr N}_{ab}$ is a matrix enforcing the GSO projection and is subject to the conditions
\begin{equation}
{\mathscr N} = T^\dagger_\text{R} \, {\mathscr N} \, T_\text{L} \,,\qquad {\mathscr N}= S^\dagger_\text{R} \, {\mathscr N} \, S_\text{L}\,, \label{GSOcond}
\end{equation}
where $T_\text{L,R}$ and $S_\text{L,R}$ represent the action of the generators of $\text{SL} (2;\mathbb{Z})$ on the characters $\chi_a$ and $\bar\chi_a$, respectively. Notice that, since any element of the modular group can be decomposed in terms of $T$ and $S$, the previous conditions also imply
\begin{equation}
{\mathscr N} = M^\dagger_\text{R} \, {\mathscr N} \, M_\text{L}\,,  \label{modinvGSO}
\end{equation} 
where $M_\text{L,R}$ are potentially different representations of the same $\text{SL} (2;\mathbb{Z})$ transformation acting on the space of characters.

In string theory, the characters $\chi_a$ of the RCFT  always come together with the contribution of the non-compact bosons, and it is natural to define the {\em dressed pseudo-characters} $\hat\chi_a = \eta^{2-D}\, \chi_a$. Strictly speaking, these are no longer characters of a RCFT, since they carry the non-trivial modular weight $1-D/2$.
As a result,  the generators of the modular group involve extra phases, and act as
\begin{equation}
\begin{split}
T:&\quad  \hat\chi_a \to \hat T_{ab} \, \hat \chi_b = e^{-2\pi i (D-2)/24}\, \frac{T_{ab}  \chi_b}{\eta^{D-2}}  \,, 
\\
S:&\quad \hat\chi_a \to \tau^{1-D/2}\, \hat S_{ab}\hat\chi_b = \tau^{1-D/2}\, i^{D/2-1}\, \frac{S_{ab}  \chi_b}{\eta^{D-2}}\,.
\end{split}\label{STmulti}
\end{equation}

The pseudo-characters $\hat\chi_a$,  from now on simply referred to as the characters $\chi_a$ with the hat omitted, admit a power series expansion in the {\em nome} $q=e^{2\pi i \tau}$ of the form
\begin{equation}
\chi_a (q) = \sum_{n=0}^\infty d_a (n) q^{H_a + n}\,,
\end{equation}
with $H_a = h_a - c/24$ expressed in terms of the conformal weight $h_a$ and the central charge $c$ of the full theory, including the $D-2$ non-compact bosons. As a result,  the 
total net number of the physical degrees of freedom at mass level $n$ reads
\begin{equation}
 d (n)  = \sum_{a,b} {\mathscr N}_{ab}\,  \bar d_a (n+ H_b -\bar H_a) \, d_b (n) \label{sasum}\,,
\end{equation}
where we have imposed the level-matching condition.

The coefficients $d_a (n)$, counting the degeneracies of states of $\chi_a$,  can be computed using Cauchy theorem,
\begin{equation}
d_a (n) = \frac{1}{2\pi i} \oint_\Gamma dq \, \frac{\chi_a (q)}{q^{n +H_a+1}} \,, \label{CMHRR}
\end{equation}
where  $\Gamma$ is any closed contour lying entirely inside the unit disk in the  complex $q$-plane and containing the origin $q=0$. 
The large-$n$ behaviour of this integral was extracted in \cite{Kani:1989im} using the circle method of Hardy and Ramanujan \cite{Hardy}, however an exact formula for the mass degeneracies, more suited for our discussion and derived in Appendix \ref{appRadexp}, can be obtained using the Rademacher \cite{Rademacher:1937a,Rademacher:1937b,Rademacher:1938} improvement on the Hardy-Ramanujan method. 

The non-vanishing contributions to the integral \eqref{CMHRR} come from the essential singularities in the $\tau$-plane located at the rational points. The {\em circle method} allows to properly sum these contributions by deforming the contour 
\begin{equation}
\Gamma \to \bigcup_{\frac{p}{\ell}\in {\mathscr F}_N} \, \gamma (p,\ell )
\end{equation}
where $\gamma (p,\ell )$ are arcs in the $\tau$-plane determined by the Ford circles tangent to the rational point $p/\ell$ and to its immediate predecessor and successor in the Farey fractions ${\mathscr F}_N$,
\begin{equation}
d_a (n) = \lim_{N\to \infty}\sum_{\ell=1}^N \sum_{{p=0 \atop (p,\ell )=1}}^{\ell-1} \int_{\gamma (p,\ell )}\, \frac{dq}{2\pi i} \frac{\chi_a (q)}{q^{n+H_a +1}}\,.
\end{equation}
In the limit $N\to \infty$ this integral {\em localises} on the essential singularities $\tau=\frac{p}{\ell} \in \mathbb{Q}$, and the behaviour of the characters $\chi_a (q)$ at such points can be connected to its behaviour $\chi_a \sim d_a (0)\, q^{H_a}$ at the cusp $q=0$ via the modular transformations 
\begin{equation}
\gamma_{\ell , p} = \begin{pmatrix} -p' & \frac{1+pp'}{\ell} \\ -\ell & p\end{pmatrix}\,. \label{gmodtr}
\end{equation}
Following the steps in \cite{Rademacher:1937a,Rademacher:1937b,Rademacher:1938}, after suitable changes of variables and deformations of contours, one can  cast the degeneracies as
\begin{equation}
d_a (n) =\lim_{N\to \infty} \,\left[\sum_{\ell =1}^N \sum_{b=0}^{M-1}\, Q^{(\ell , n )}_{ab} \, f_{b} (\ell , n) + S_N \right] \,.
\end{equation}
The matrix
\begin{equation}
Q^{(\ell,n)}_{ab} =  i^{1-D/2}\, \sum_{p=0\atop  (p,\ell)=1}^{\ell-1} \left( M^{-1}_{\ell,p} \right)_{ab} \, e^{-\frac{2\pi i}{\ell} (p  (n+H_a ) - p' H_b )} \label{Qfunct}
\end{equation}
is expressed in terms of the representation $M_{\ell,p}$ of the modular transformation \eqref{gmodtr} on the characters, $\chi_a \to (-\ell \, \tau + p)^{1-D/2}\, (M_{\ell , p})_{ab} \chi_b$, while 
\begin{equation}
f_{b} (\ell ,n) = - i \, \frac{d_b (0)}{\ell}\, \left(\frac{|H_b|}{n+H_a}\right)^{D/4} \, \int_{1-i\infty}^{1+i\infty} dw\, w^{D/2-1}\, e^{\frac{2\pi}{\ell} \sqrt{|H_b | \, (n+H_a)} \left( \frac{1}{w} - \text{sgn} (H_b)\,  w \right)} \,. \label{integral}
\end{equation}
Finally, $S_N$ collects the error terms associated to the exact behaviour of the characters at the cusp and the approximation in the definition of the contours, and it can be shown to vanish in the $N\to \infty$ limit. The integral \eqref{integral} can be evaluated using Jordan's lemma and yields the non-vanishing result 
\begin{equation}
f_{b} (\ell , n) = \frac{2\pi d_b (0)}{\ell}\, \left(\frac{|H_b|}{n+H_a}\right)^{D/4} \, I_{D/2} \left( \frac{4\pi}{\ell} \sqrt{ |H_b | (n+ H_a )}\right)\,,
\end{equation}
only when $\text{sgn} (H_b ) =-1$, with $I_\alpha (z)$ the modified Bessel function of the first kind. As a result, the {\em exact} mass degeneracy reads
\begin{equation}
d_a (n) =\sum_{b\, |\, H_b <0} \sum_{\ell =1}^\infty  \, Q^{(\ell , n )}_{ab} \,  f_{b}  (\ell , n) \,,
\end{equation}
and receives contributions only from the tachyonic characters with $H_b <0$, which are always present in the CFT of any string construction, although they may be eliminated from the physical spectrum by the GSO projection.

The asymptotic expansion
\begin{equation}
 I_\alpha (x) \sim \frac{e^x}{\sqrt{2\pi x}} \, \left( 1- \frac{4\alpha^2-1}{8x} + \ldots \right)\,,
\end{equation}
valid for positive $x$, determines the large-mass behaviour of the degeneracies
\begin{equation}
d_a (n) \sim \sum_{b \, |\, H_b <0} \frac{d_b (0)}{\sqrt{2}}\, \frac{|H_b|^{(D-1)/4}}{n^{(D+1)/4}}\,
\left[ Q^{(1,n)}_{ab} e^{4\pi \sqrt{|H_b| n}} + \frac{Q^{(2,n)}_{ab}}{\sqrt{2}}\, e^{2\pi \sqrt{|H_b| n}}  + \frac{Q^{(3,n)}_{ab}}{\sqrt{3}}\,  e^{\frac{4\pi }{3}\sqrt{|H_b| n}} +\ldots \right]\,,
\end{equation}
with $Q_{ab}^{(1,n)} = i^{1-D/2}\, S_{ab}$ given by the $S$ matrix. The deepest tachyon, with the smallest conformal weight, included in $\chi_0$, is associated to the ubiquitous NS vacuum and plays the role of the identity in the RCFT. As a result,  it dictates the leading growth
\begin{equation}
d_a (n) \sim e^{4\pi \sqrt{c n/24}} + \ldots \,, \label{leadgrw}
\end{equation}
in agreement with the Cardy formula \cite{Cardy:1981fd}, and this contribution is universal since $S_{a0} = i^{D/2-1}/\sqrt{M}$. 
The sub-leading exponentials depend instead on the data of the RCFT. 

\section{A Proof of Misaligned Supersymmetry}\label{main}

We have now all the ingredients to connect the large-$n$ behaviour of $d(n)$ with the classical stability of a string vacuum. To this end, following \cite{Dienes:1994np}, we continue the integer $n$ to the reals, introduce the enveloping functions $d_a (n) \to \Phi_a (n)$, and construct the sector averaged sum
\begin{equation}
\langle d (n) \rangle = \sum_{a,b} {\mathscr N}_{ab}\, \bar\Phi_a (n+ H_b -\bar H_a) \Phi_b (n )\,,
\end{equation}
 whose asymptotic behaviour dictates the presence/absence of physical tachyons. 
 
From eqs \eqref{sasum} and \eqref{leadgrw} it follows that the $\text{SL} (2;\mathbb{C})$ invariant vacuum in principle determines the leading large-$n$ behaviour, 
\begin{equation}
\langle d (n) \rangle \sim  \, e^{4\pi \sqrt{n} \left( \sqrt{c_\text{L}/24}+\sqrt{ c_\text{R} /24}\right)}\, \sum_{a,b} {\mathscr N}_{ab} = e^{C_\text{tot} \, \sqrt{n}} \, \, \sum_{a,b} {\mathscr N}_{ab} \,.
\end{equation}
However, whether the asymptotic growth is controlled by $C_\text{tot}$ or not depends entirely on the properties of ${\mathscr N}$. Recall that the GSO matrix must yield a modular invariant partition function, and is thus subject to the conditions \eqref{GSOcond}. As a result,
\begin{equation}
{\mathscr N}_{00} = \frac{1}{M} \, \sum_{a,b} {\mathscr N}_{ab}
\end{equation}
since, for a RCFT, where all  $M$ characters have been resolved, $S_{a0}=i^{D-2-1}/\sqrt{M}$. Moreover, the entries ${\mathscr N}_{ab}$ can only be $\pm 1$ or zero, which yield to the following two scenarios:
\begin{enumerate}
\item If the spectrum {\em does} contain the leading tachyon, {\em i.e.} ${\mathscr N}_{00}=1$, one has 
\begin{equation}
\sum_{a,b} {\mathscr N}_{ab} \neq 0 \qquad \text{and} \qquad \langle d (n) \rangle \sim e^{C_\text{tot} \, \sqrt{n}}\,.
\end{equation}
\item If the spectrum {\em does not} contain the leading tachyon, {\em i.e.} ${\mathscr N}_{00}=0$,  then
\begin{equation}
\sum_{a,b} {\mathscr N}_{ab} = 0 \qquad \text{and} \qquad \langle d (n) \rangle \sim e^{C_\text{eff} \, \sqrt{n}}\label{FBNzero}
\end{equation}
with an {\em effective central charge} $C_\text{eff} < C_\text{tot}$. 
\end{enumerate}
The condition ${\mathscr N}_{00}=0$ clearly requires that fermions be present in the spectrum and, in particular the number of bosonic sectors must equal the number of fermionic ones. On the contrary, in the absence of extended symmetries, namely when each character appears only once in  \eqref{partition}, the condition ${\mathscr N}_{00}=1$ implies that the spectrum only contains bosonic excitations. These observations were already contained in \cite{Dienes:1994np} and, in fact, represent the main result of Dienes' paper. However, it is important to stress that $C_\text{eff} < C_\text{tot}$ does {\em not} imply classical stability, as sometimes stated in the literature, since, although it is true that the leading tachyon must be absent, the condition ${\mathscr N}_{00} =0$ does not automatically exclude the possibility that other tachyons be present. Indeed, as we shall see in the following Sections, this is precisely what happens in most non-supersymmetric theories. 

The analysis of the sub-dominant contributions to $\langle d (n) \rangle$ is a bit more involved. To start, we notice that the $Q$'s are periodic functions of $n$, with period $\ell$, 
\begin{equation}
Q^{(\ell,n+\ell m)}_{ab} = Q^{(\ell,n)}_{ab}\,,
\end{equation}
for any integer $m$.  Moreover, since the number of physical states \eqref{sasum} depends on the product of the $Q$ functions from the holomorphic and anti-holomorphic characters,  it is convenient to decompose the degrees of freedom into classes organised by the common periodicity $v_{\ell,\bar \ell} = \ell\, \bar \ell /\text{gcd} (\ell,\bar \ell)$ of the two $Q$'s and associate to each class its own enveloping function $\Phi_a (n , \{ w\} )$, as in \cite{Cribiori:2020sct}. As a result,
\begin{equation}
\begin{split}
\langle d (n ) \rangle &= \sum_{a,b}\sum_{\{w  \}} {\mathscr N}_{ab} \, \bar \Phi_a (n, \{ w  + H_b - \bar H_a \} )\, \Phi_b (n, \{ w  \} )
\\
&\equiv \sum_{a,b} \sum_{{c,d \atop \, H_c, \bar H_d <0}}  \sum_{\ell,\bar \ell =1}^{\infty} \sum_{w=0}^{v_{\ell,\bar \ell}-1} {\mathscr N}_{ab}\, Q^{(\ell,w)}_{bc}\,  \bar Q^{(\bar \ell ,  w+H_b - \bar H_a)}_{ad} \, f_c (\ell,n )\,  \bar f_d (\bar \ell , n) \,.
\end{split}
\end{equation}
Moreover, since one can always write $w=k_\ell + \ell r$, and 
\begin{equation}
\sum_{w=0}^{v_{\ell,\bar \ell} -1} = \sum_{k_\ell =0}^{\ell-1} \sum_{r=0}^{\frac{\bar \ell}{\text{gcd} (\ell,\bar \ell)}-1} \,,
\end{equation}
the periodicity of the $Q$'s can be exploited to cast the sector averaged sum as
\begin{equation}
\langle d (n) \rangle = \sum_{a,b} \sum_{{c,d \atop \, H_c, \bar H_d <0}} \sum_{\ell,\bar \ell =1}^{\infty}\sum_{k_\ell=0}^{\ell-1} \sum_{r=0}^{\frac{\bar \ell}{\text{gcd} (\ell,\bar \ell)}-1} {\mathscr N}_{ab} \, Q^{(\ell,k_\ell )}_{bc}\,  \bar Q^{(\bar \ell ,  k_\ell + r \ell + H_b - \bar H_a)}_{ad} \, f_c (\ell,n )\,  \bar f_d (\bar \ell , n) \,.\label{maineq}
\end{equation}
This is the expression from which we can extract our main results. We have to distinguish the two cases $\ell = \bar \ell$ and $\ell\neq \bar \ell$. In the latter case, if $\ell$ does not divide $\bar\ell$, there is no contribution to the sector averaged sum, since
\begin{equation}
\sum_{r=0}^{\frac{\bar \ell}{\text{gcd} (\ell,\bar \ell)}-1} \bar Q^{(\bar \ell ,  k_\ell + r \ell + H_b - \bar H_a)}_{ad} =0\,, \label{eqllbdiff}
\end{equation}
which follows from 
\begin{equation}
\begin{split}
\sum_{r=0}^{\frac{\bar \ell}{\text{gcd} (\ell,\bar \ell)}-1} \bar Q^{(\bar \ell ,  k + r \ell )}_{ad} &= \sum_{r=0}^{\frac{\bar \ell}{\text{gcd} (\ell,\bar \ell)}-1} \sum_{\bar p=0\atop (\bar p,\bar \ell )=1}^{\bar \ell -1} e^{-\frac{2 \pi i}{\bar \ell} (\bar p ( k + r \ell)+  \bar p ' \bar H_d )}\, \left( M^{-1}_{\bar \ell , \bar p}\right)_{ad}
\\
&=\sum_{\bar p=0\atop  (\bar p, \bar \ell)=1}^{\bar \ell -1}  e^{-\frac{2 \pi i}{\bar \ell} (\bar p k +  \bar p ' \bar H_d )}\, \left( M^{-1}_{\bar \ell , \bar p}\right)_{ad}\, \frac{1-e^{-2\pi i \bar p \frac{\bar \ell}{\text{gcd} (\ell,\bar \ell)}}}{1-e^{- 2\pi i \frac{\bar p \ell}{\bar \ell}}} 
\\
&=0 \,,
\end{split}
\end{equation}
since $\bar \ell$ is an integer multiple of $\text{gcd} (\ell, \bar \ell)$. Similarly, if $\bar\ell = m \ell$, for some integer $m$, eq. \eqref{eqllbdiff} still holds since $\bar p$ and $\bar\ell$ must be co-prime, and thus
\begin{equation}
\sum_{r=0}^{m-1} e^{-2 \pi i \bar p r/m} = 0\,.
\end{equation}

The case $\ell=\bar \ell$ is a bit more involved and, as we shall see, discriminates between tachyonic and non-tachyonic string vacua. The $Q$ functions have now the same periodicity, which implies that in eq. \eqref{maineq} the sum over $r$ is trivial, since nothing depends on $r$, and
\begin{equation}
\begin{split}
\langle d (n) \rangle &= \sum_{a,b} \sum_{{c,d \atop \, H_c, \bar H_d <0}} \sum_{\ell =1}^{\infty} \sum_{ p=0\atop (p, \ell )=1}^{ \ell -1} \sum_{\bar p=0\atop (\bar p,  \ell )=1}^{ \ell -1} {\mathscr N}_{ab} \, \left( M^{-1}_{\ell ,  p}\right)_{bc} \, \left( M^{-1}_{ \ell , \bar p}\right)_{ad}^*
\, e^{-\frac{2\pi i}{\ell}  [ (p-\bar p )  H_b - (p' H_c - \bar p ' \bar H_d )]}\,
 f_c (\ell,n )\,  \bar f_d ( \ell , n)  
 \\
 &\qquad \times \sum_{k_\ell=0}^{\ell-1} e^{- \frac{2 \pi i}{\ell} (p-\bar p ) k_\ell }  
 \,.
 \end{split}
\end{equation}
The sum over $k_\ell$ imposes the condition $p=\bar p$ which identifies the holomorphic and anti-holomor\-phic modular transformations. For large $n$, using the condition \eqref{modinvGSO} of modular invariance on the GSO matrix ${\mathscr N}_{ab}$,  and taking into account that in string theory tachyons have $-1\le \bar H_a , H_b <0$, one arrives at the final result
\begin{equation}
\langle d(n) \rangle = \sum_{\ell=1}^{\infty} 
\sum_{{a, b\atop \bar H_a = H_b <0}}  \,\ell\, \varphi (\ell ) \, {\mathscr N}_{ab}\,  f_b (\ell,n )\,  \bar f_a ( \ell , n)  \,,
 \end{equation}
 where $\varphi (\ell)$ is the Euler totient function.   {\em The exponential growth of the sector averaged sum is thus directly linked to the presence of tachyons in the physical string spectrum.} In fact,  if physical tachyons, with $H_b = \bar H_a <0$, are present in the string spectrum, then
 \begin{equation}
 \langle d(n) \rangle \sim \sum_{a,b \atop H_b = \bar H_a <0} {\mathscr N}_{ab}\,  d_b (0) \, \bar d _a (0)\, \frac{|H_b|^{(d-1)/2}}{2\, n^{(d+1)/2}}\, 
 \sum_{\ell=1}^{\infty} \, \varphi (\ell) \, e^{\frac{8\pi}{\ell} \sqrt{|H_b|\, n}} \,.
 \end{equation}
The exponential growth is then dictated by the conformal weight $H_b$ of the {\em deepest} physical tachyon,  $C_\text{eff} = 8 \pi \sqrt{|H_b|}$. Instead, in the absence of on-shell tachyons the sector averaged sum $\langle d (n) \rangle$ vanishes, and thus
\begin{equation}
C_\text{eff} =0\,.
\end{equation}
 
 To reiterate, we have proven that the asymptotic growth rate of the sector averaged sum is dictated by the mass of the {\em lightest} states, whether tachyonic or massless,
\begin{equation}
C_\text{eff} = 4 \pi \sqrt{|\alpha' m^2_\text{lightest} |} \le C_\text{tot}\,,
\end{equation}
and thus the {\em necessary and sufficient} condition for classical stability is the vanishing of both the sector averaged sum and the effective central charge, as conjectured by Dienes in \cite{Dienes:1994np}. Notice that the deepest tachyon determines also the {\em effective} Hagedorn temperature $T_\text{eff}$ of strings at finite temperature \cite{Dienes:2012dc}, and thus links it to $C_\text{eff}$. 

Our analysis is fully general and applies to any vacuum of oriented closed strings. It extends the discussion of \cite{Cribiori:2020sct} which heavily relies on the representation of the characters in terms of eta quotients of special type, which is a rather restrictive requirement, not met by most non-supersymme\-tric string vacua.

\section{Non-Supersymmetric Heterotic Vacua in $D=10$} \label{10d}

A simple arena where to test our result is ten-dimensional closed-string vacua with no space-time supersymmetry. Indeed, one-loop modular invariance allows for many consistent constructions in ten dimensions, most of which do not enjoy space-time supersymmetry. These non-supersymmetric vacua can be divided into  three different classes: tachyonic theories with only bosonic excitations, tachyonic theories with both fermionic and bosonic fields, and a single theory with no tachyons. The unique representative of the last class is the celebrated $\text{SO}(16) \times \text{SO}(16)$ heterotic theory \cite{Dixon:1986iz, Alvarez-Gaume:1986ghj} with partition function
\begin{equation}
\begin{split}
{\mathscr Z}_{16} &= O_8 \, (\bar V_{16} \bar C_{16} + \bar C_{16} \bar V_{16} ) + V_8 (\bar O_{16} \bar O_{16} + \bar S_{16} \bar S_{16} ) 
\\
&\qquad - S_8 \, (\bar O_{16} \bar S_{16} + \bar S_{16} \bar O_{16} )   - C_8 \, ( \bar V_{16} \bar V_{16} + \bar C_{16} \bar C_{16} )\,,
\end{split}\label{HetO16O16}
\end{equation}
while the first class comprises the type 0A and 0B strings \cite{Dixon:1986iz}, with partition functions
\begin{equation}
{\mathscr Z}_{0\text{A}} = |O_8 |^2 + |V_8 |^2 + S_8 \bar C_8 + C_8 \bar S_8 \,, \qquad {\mathscr Z}_{0\text{B}} = |O_8 |^2 + |V_8|^2 + |S_8 |^2 + |C_8|^2\,. 
\label{0AB}
\end{equation}
The second class is richer and contains five heterotic vacua with gauge groups $\text{SO} (32)$, $\text{SO}(16) \times \text{E}_8$, $\text{SO}(8) \times \text{SO}(24)$, $(\text{E}_7 \times \text{SU}(2) )^2$ and $\text{SU}(16)$ \cite{Dixon:1986iz}. These theories present a similar behaviour and, for simplicity,  we shall concentrate on the $\text{SO}(32)$ theory with partition function
\begin{equation}
{\mathscr Z}_{32} = O_8 \, \bar V_{32} + V_8 \, \bar O_{32}  - S_8 \, \bar S_{32} - C_8 \, \bar C_{32}\,. 
\label{HetO32}
\end{equation}
Notice that in writing eqs. \eqref{HetO16O16}, \eqref{0AB} and \eqref{HetO32} we have used the $\text{SO} (2n)$ characters, as defined in \cite{Angelantonj:2002ct}.

As anticipated in the previous Section, strictly speaking these string theories do not correspond to RCFTs because of the presence of non-compact bosons. Still, following \cite{Kani:1989im}, we can overcome this problem by defining the  pseudo-characters
\begin{equation}
(O_{2n} , V_{2n} , S_{2n} , C_{2n} ) \to \left(\frac{O_{2n} }{\eta^8}, \frac{V_{2n}}{\eta^8} , \frac{S_{2n}}{\eta^8} , \frac{C_{2n}}{\eta^8}\right)\,,
\end{equation}
and including suitable phases in the modular transformations 
\begin{equation}
T \, :\quad   (O_{2n} , V_{2n} , S_{2n} , C_{2n} ) \to e^{- i \pi (n+8)/12}\,  (O_{2n} , - V_{2n} , e^{i \pi n/4} \, S_{2n} , e^{i \pi n/4} \, C_{2n} ) \,, \label{TSO2n}
\end{equation}
and 
\begin{equation}
S\,:\quad \begin{pmatrix} O_{2n} \\ V_{2n} \\ S_{2n} \\ C_{2n}\end{pmatrix} \to \tau^{-4} \, \frac{1}{2}\begin{pmatrix} 1 & 1 & 1 & 1 \\ 1 & 1 & -1 & -1 \\ 1 & -1 & i^{-n} & - i^{-n}\\ 1 & -1 & - i^{-n} &  i^{-n} \end{pmatrix} \, \begin{pmatrix} O_{2n} \\ V_{2n} \\ S_{2n} \\ C_{2n}\end{pmatrix}\,. \label{SSO2n}
\end{equation}

We can discuss all these theories at once by noticing that their partition functions can be compactly written as
\begin{equation}
{\mathscr Z}_A = \sum_{a,b=0}^3 \bar R^A_a\, {\mathscr N}_{ab} \, L_b \,,
\end{equation}
where $L = (O_8 , V_8 , S_8 , C_8)$ denotes the left-moving characters, which are common to all ten-dimen\-sional non-supersymmetric theories, and $\bar R_a^A$ denotes the right-moving characters, which depend on the specific model $A=16,0,32$, and can be extracted from eqs. \eqref{HetO16O16}, \eqref{0AB} and \eqref{HetO32}, respectively. The corresponding GSO matrices read
\begin{equation}
{\mathscr N}_{0A} = \begin{pmatrix} 1 & 0 & 0 & 0 \\ 0 & 1 & 0 & 0 \\ 0 & 0 & 0 & 1 \\ 0 & 0 & 1 & 0 \end{pmatrix}\,,
\qquad 
{\mathscr N}_{0B} = \begin{pmatrix} 1 & 0 & 0 & 0 \\ 0 & 1 & 0 & 0 \\ 0 & 0 & 1 & 0 \\ 0 & 0 & 0 & 1 \end{pmatrix}\,,
\end{equation}
for the type 0A and type 0B theories, while
\begin{equation}
{\mathscr N}_\text{het} = \begin{pmatrix} 0 & 1 & 0 & 0 \\ 1 & 0 & 0 & 0 \\ 0 & 0 & -1 & 0 \\ 0 & 0 & 0 &-1 \end{pmatrix}\,,
\end{equation}
for the heterotic theories. 

The universal holomorphic characters $L_a$ have  $H_a = (-\frac{1}{2} , 0 ,0 ,0)$, and using the explicit form \eqref{TSO2n} and \eqref{SSO2n} of the $T$ and $S$ modular matrices, one finds
\begin{equation}
\begin{split}
\Phi_a (n ; \{ w_\ell \})  &= \frac{1}{ 2^{\frac{15}{4}}\, n^\frac{11}{4}}\, \left[ e^{4\pi \sqrt{n/2 }} - \sqrt{2}\,\delta_{a0}\,  (-1)^{w_2} \, e^{2\pi \sqrt{n/2}} \right.
\\
& \qquad \qquad \left. +\tfrac{2}{\sqrt{3}}\, \cos\left[ \tfrac{2}{3} \pi (w_3 + (-1)^{\delta_{a0}}) \right] \, e^{\frac{4\pi}{3} \sqrt{n/2}} +\ldots \right] \,,
\end{split}
\end{equation}
for the asymptotic behaviour of the degeneracies, where the $w_\ell$ take integer values in the interval $0, \ldots , \ell-1$. Notice that $L_0 = O_8$  is the unique tachyonic character, associated to the NS vacuum, which reflects the presence of a single {\em leading} exponential growth $e^{4\pi \sqrt{n/2}}$. 

As for the anti-holomorphic characters $\bar R_a^A$, they carry weights
\begin{equation}
\begin{array}{ll}
\bar H_a= (-1, \frac{1}{2} , 0,0) &  \text{for the } \text{SO}(16) \times \text{SO}(16) \text{ heterotic theory}
\\
\bar H_a= (-\frac{1}{2},0,0,0) &  \text{for the 0A and  0B  theories}
\\
\bar H_a= (-1,-\frac{1}{2}  , 1, 1) & \text{for the } \text{SO}(32) \text{ heterotic theory}
\end{array}
\end{equation}
and using the explicit expression for the $T$ and $S$ modular matrices, which can be read/derived from eqs. \eqref{TSO2n} and \eqref{SSO2n}, one finds
\begin{equation}
\begin{split}
\bar \Phi_a ^{(16)} (n ; \{ \bar w_l \} ) &=\frac{1}{2^{\frac{3}{2}}\, n^{\frac{11}{4}}} \left[ e^{4\pi \sqrt{n}} + \sqrt{2} \, \delta_{a1}\, (-1)^{\bar w_2} \, e^{2\pi \sqrt{n}} \right.
\\
& \qquad \qquad\left. +\tfrac{2}{\sqrt{3}} \, \cos \left[ \tfrac{2}{3} \pi (\bar w_3 + \delta_{a0}) \right]\, e^{\frac{4\pi}{3}\sqrt{n}}+\ldots \right]\,,
\end{split}
\end{equation}
for the $\text{SO}(16) \times \text{SO}(16)$ theory,
\begin{equation}
\begin{split}
\bar \Phi_a ^{(32)} (n , \{ \bar w_\ell \} ) &=\frac{1}{2^{\frac{3}{2}}\, n^{\frac{11}{4}}} \left[ e^{4\pi \sqrt{n}} + 2^{\frac{11}{4}}\, (-1)^{\delta_{a2}+\delta_{a3}}\, e^{4\pi \sqrt{n/2}}  \right.
\\
&\qquad \qquad +\sqrt{2}\, \delta_{a1}\,  (-1)^{\bar w_2}\, e^{2\pi \sqrt{n}} - 2^{\frac{13}{4}}\, \delta_{a0}\, (-1)^{\bar w_2} \,   e^{2\pi \sqrt{n/2}}
\\
& \qquad \qquad +\tfrac{2}{\sqrt{3}}\, \cos \left[ \tfrac{2}{3} \pi (\bar w_3 - \delta_{a1} ) \right]\, e^{\frac{4\pi}{3} \sqrt{n}}
\\
&\qquad \qquad \left. - \tfrac{ 2^{\frac{15}{4} }}{\sqrt{3}}\, (-1)^{\delta_{a1}}\, \cos \left[ \tfrac{2}{3}\pi (\bar w_3 + 1 -\delta_{a1})\right]\, e^{\frac{4\pi}{3} \sqrt{n/2}} 
+\ldots \right]\,,
\end{split}
\end{equation}
for the $\text{SO}(32)$ theory, while $\bar \Phi_a^{(0)} (n,\{ \bar w_\ell \} ) = \Phi_a  (n,\{ \bar w_\ell \})$ for the 0A and 0B theories. Notice, that in $\bar \Phi_a ^{(32)} (n , \{ \bar w_\ell \} )$ there are two {\em leading} exponential growths $e^{4\pi \sqrt{n}}$ and $e^{4\pi \sqrt{n/2}}$ associated, respectively, to the two tachyonic characters $\bar R_0$ and $\bar R_1$, with $\bar H_0= -1$ and $\bar H_1 = -\frac{1}{2}$.

The sector averaged sum  is then given by
\begin{equation}
\langle d(n) \rangle_A =\sum_{a,b=0}^3\, \sum_{\{w_\ell\} } {\mathscr N}_{ab}\, \bar \Phi_a^{(A)} (n,\{w_\ell + H_b - \bar H_a \} ) \, \Phi_b (n,\{ w_\ell\} ) \,,
\end{equation}
where the choice $\bar w_\ell = w_\ell + H_b - \bar H_a$ follows from the level matching condition, and the sum is extended over all integers $w_\ell = 0, \ldots , \ell-1$, for each $\ell=2,3,\ldots$.  It is straightforward to see that the growth of  $\langle d(n) \rangle_0 \sim e^{4\pi \sqrt{2 n} }$ for the type 0A and 0B theories is dictated by the total central charge $C_\text{tot} = 4 \pi  \sqrt{2} $, while for the heterotic models the leading term $e^{4 \pi (1+1/\sqrt{2}) \sqrt{ n}}$ cancels upon summing over the sectors $a, b=0,\ldots, 3$. This reflects the fact that the type 0 theories only have bosonic excitations while the heterotic ones have both bosons and fermions in their spectrum, and $\sum_{ab} {\mathscr N}_{ab} =0$. Moreover, while for the tachyonic $\text{SO}(32)$ theory
\begin{equation}
\langle d(n) \rangle_{32} = \frac{1}{\sqrt{2}\, n^{\frac{11}{2}}} \left[ e^{4 \pi \sqrt{2n}} + e^{2 \pi \sqrt{2n}} + 2 e^{\frac{2\pi}{3} \sqrt{2n}}+\ldots  \right]\,,
\end{equation}
and $C_\text{eff} = 4 \pi \sqrt{2} < C_\text{tot}$, for the non-tachyonic $\text{SO}(16) \times \text{SO}(16)$ model $C_\text{eff} =0$ since the GSO projection eliminates the tachyons from the physical spectrum, and as a result $\langle d( n) \rangle =0$. This is in agreement with our general result given in the previous Section and with the result of \cite{Cribiori:2020sct} for the $\text{SO}(16) \times \text{SO}(16)$ theory.

\section{Scherk-Schwarz Reductions at Rational Points} \label{ScSc}

A more interesting class of non-supersymmetric vacua can be constructed in lower dimensions by employing the Scherk-Schwarz mechanism \cite{Scherk:1979zr,Ferrara:1987es}. This can be conveniently realised as a freely acting orbifold \cite{Kounnas:1989dk}, where the action of a supersymmetry breaking generator $g$ is combined with a suitable shift $\delta$ along compact directions. In its simplest nine-dimensional incarnation $g = (-1)^F$, with $F$ the space-time fermion number, while $\delta$ acts as $ y \to y +\pi R$ on the compact    coordinate $y$ parametrising the circle $S^1 (R)$ of radius $R$. In the heterotic case, one can consider more general orbifolds where the space-time fermion number is combined with an action on the gauge degrees of freedom, in accordance with modular invariance.  In this Section, however, we shall consider the Scherk-Schwarz reduction of the type IIB superstring since it shares the main features with any generic construction, with the advantage of being quite simple. The torus partition function 
\begin{equation}
\begin{split}
{\mathscr Z} &= \tfrac{1}{2} |V_8 - S_8|^2 \, \sum_{m,n} \Lambda_{m,n} (R) + \tfrac{1}{2} |V_8 + S_8 |^2 \, \sum_{m,n} (-1)^m \, \Lambda_{m,n} (R)
\\
& \qquad + \tfrac{1}{2} |O_8 - C_8|^2 \, \sum_{m,n} \Lambda_{m,n+\frac{1}{2}} (R) + \tfrac{1}{2} |O_8 + C_8|^2 \, \sum_{m,n} (-1)^m\, \Lambda_{m,n+\frac{1}{2}} (R)  \,,
\end{split}
\end{equation}
clearly exhibits in the first line the action of the $(-1)^F\, \delta$ generator on the original spectrum of the IIB superstring, while the second line, involving the flipped GSO projection, is required by modular invariance. The Kaluza-Klein momenta and windings associated to the compact direction contribute with the standard Narain lattice
\begin{equation}
\Lambda_{m,n} = \frac{q^{\frac{\alpha'}{4} \left( \frac{m}{R} + \frac{nR}{\alpha '}\right)^2}}{\eta} \, \frac{\bar q^{\frac{\alpha'}{4} \left( \frac{m}{R} - \frac{nR}{\alpha '}\right)^2}}{\bar \eta}\,.\label{NarainL}
\end{equation}
In the decompactification limit, $R\to \infty$, the orbifold action  is trivialised and one recovers the supersymmetric  IIB theory, while for generic values of the radius $R$ supersymmetry is spontaneously broken and the gravitini acquire a mass $m\simeq 1/R$. The excitations of the NS-NS vacuum in $| O_8|^2$ now survive the GSO projection in the twisted sector, and the lightest state has mass
\begin{equation}
m^2_{|O_8|^2} =-\frac{1}{2 \alpha'} +\frac{1 }{4} \left( \frac{R}{2\alpha '}\right)^2\,. \label{masstachyon}
\end{equation}
This scalar is then massive for large values of $R$, but turns tachyonic below the critical radius $R_c = 2 \sqrt{2\alpha '}$. As we decrease the radius, more and more states become tachyonic, and it is then clear that these models represent an ideal ground to study the realisation of {\em misaligned supersymmetry} in string theory\footnote{See for instance \cite{Abel:2015oxa} for a first study of {\em misaligned supersymmetry} in the context of Scherk-Schwarz reductions.}. 

To illustrate the analysis of Section \ref{main} on the degeneracies of states, we need to select rational values for $ R^2/\alpha ' = s/t \in\mathbb{Q}$ since, in this case, the Narain lattice reduces to an RCFT 
\begin{equation}
\sum_{m,n} \Lambda_{m,n} (R) \to \sum_{\alpha=0}^{2st-1} \lambda_\alpha \, \bar\lambda_{\alpha l}\,, \label{NarainRCFT}
\end{equation}
with the $2st$ characters defined as
\begin{equation}
\lambda_\alpha (q) = \sum_m \frac{q^{st \left( m + \frac{\alpha}{2st}\right)^2}}{\eta (q)}\,. \label{defl}
\end{equation}
Notice, that $\lambda_0$ and $\lambda_{st}$ are real, while $\lambda_\alpha$ and $\lambda_{2st-\alpha}$, $\alpha=1,\ldots , st-1$,  form conjugate pairs. 
The $\lambda$'s have conformal weight $h_\alpha = \alpha^2/4st$, and thus $H_\alpha = h_\alpha -\frac{1}{24}$, for $\alpha =0 , \ldots , st$, with conjugate pairs  carrying the same weight. In eq. \eqref{NarainRCFT} the anti-holomorphic characters have index $\alpha l$, where $l = rt + s v$, with the integers $r$ and $v$ satisfying the relation $rt - vs =1$, and the label $\alpha l$ is defined modulo $2st$. 
As shown in the Appendix \ref{appScScrational}, these characters are eigenstates of the shift operator $\delta$ only for even $s$, but must be broken into sub-characters for odd $s$. For simplicity, here we shall restrict the discussion  to the even-$s$ case, where
\begin{equation}
\delta\ :\quad \lambda_\alpha \to (-1)^{\alpha/2t} \, \lambda_\alpha\,,
\end{equation}
and we  also take $t=1$, so that the condition $rt - vs =1$ can be easily solved by $r=1$ and $v=0$, for any $s$. Other choices for $s$ and $t$ yield equivalent results. The action of the modular group on these characters is encoded in the $T$ and $S$ matrices
\begin{equation}
T_{\alpha \beta} = e^{i \pi \left( \frac{\alpha^2}{2s}-\frac{1}{12}\right)}\, \delta_{\alpha \beta}\,,\qquad
S_{\alpha \beta} = \frac{e^{2 \pi i \frac{\alpha \beta}{2s}}}{\sqrt{2s}}\,.
\end{equation}
Taking all this into account, the torus partition function becomes
\begin{equation}
\begin{split}
{\mathscr Z} &= \sum_{a=0}^{s-1}\left(  |\chi_{2a+ 2s}|^2 + |\chi_{2a+4s}|^2 \right) 
\\
&\quad - \sum_{a=0}^{s-1} \left( \chi_{2a+1+2s} \bar \chi_{2a+1+4s} + \chi_{2a+1+4s}\, \bar\chi_{2a+1+2s} \right)
\\
&\quad +\sum_{a=0}^{\frac{s}{2}-1} \left( \chi_{2a+\sigma}\, \bar\chi_{2a+\sigma+s} + \chi_{2a+\sigma+s}\, \bar\chi _{2a+\sigma} \right.
\\
&\qquad\qquad\qquad \left.
+ \chi_{2a+\sigma+6s}\, \bar\chi_{2a+\sigma+7s} + \chi_{2a+\sigma+7s}\, \bar\chi _{2a+\sigma+6s} \right)
\\
&\quad -\sum_{a=0}^{\frac{s}{2}-1} \left( \chi_{2a+1-\sigma} \, \bar \chi_{2a+1-\sigma+7s} + \chi_{2a+1-\sigma+7s}\, \bar \chi_{2a+1-\sigma} \right.
\\
&\qquad \qquad \qquad\left.
+\chi_{2a+1-\sigma+s} \, \bar\chi_{2a+1-\sigma+6s} + \chi_{2a+1-\sigma+6s} \, \bar \chi_{2a+1-\sigma+s}
\right) \,,
\end{split}\label{ScScPart}
\end{equation}
where in the third and fourth sums one has to distinguish the two cases $s=2(2m+\sigma)$ with $\sigma=0, 1$ while, as dictated by spin-statistics, the minus signs reflect the presence of space-time fermions. The new characters are 
\begin{equation}
\{\chi_a\}_{a=0}^{8s-1} = (O_8 , V_8 ,  S_8 ,  C_8 ) \otimes \{\lambda_\alpha \}_{\alpha=0}^{2s-1}\,.
\end{equation}

The characters $\chi_a$ and $\chi_{2s-a}$ have shifted conformal weight $H_a = \frac{a^2}{4s}-\frac{1}{2}$, and therefore are tachyonic for $a <\sqrt{2s}$. However, since they appear in the partition function in the combination
\begin{equation}
\chi_{2a+\sigma}\, \bar\chi_{2a+\sigma+s} + \chi_{2a+\sigma+s}\, \bar\chi _{2a+\sigma} 
\end{equation}
the only states which are level-matched are 
\begin{equation}
\chi_{s/2} \, \bar \chi_{3s/2} + \chi_{3s/2} \, \bar\chi_{s/2}\,,
\end{equation}
which are tachyonic for $s < 8$. This agrees with the result of eq. \eqref{masstachyon} valid at irrational values of $R$, since  now $R^2 = s \, \alpha'$. 

\begin{figure}[h!]
	\centering
	\begin{subfigure}[b]{0.4\linewidth}
	\includegraphics[width=\linewidth]{"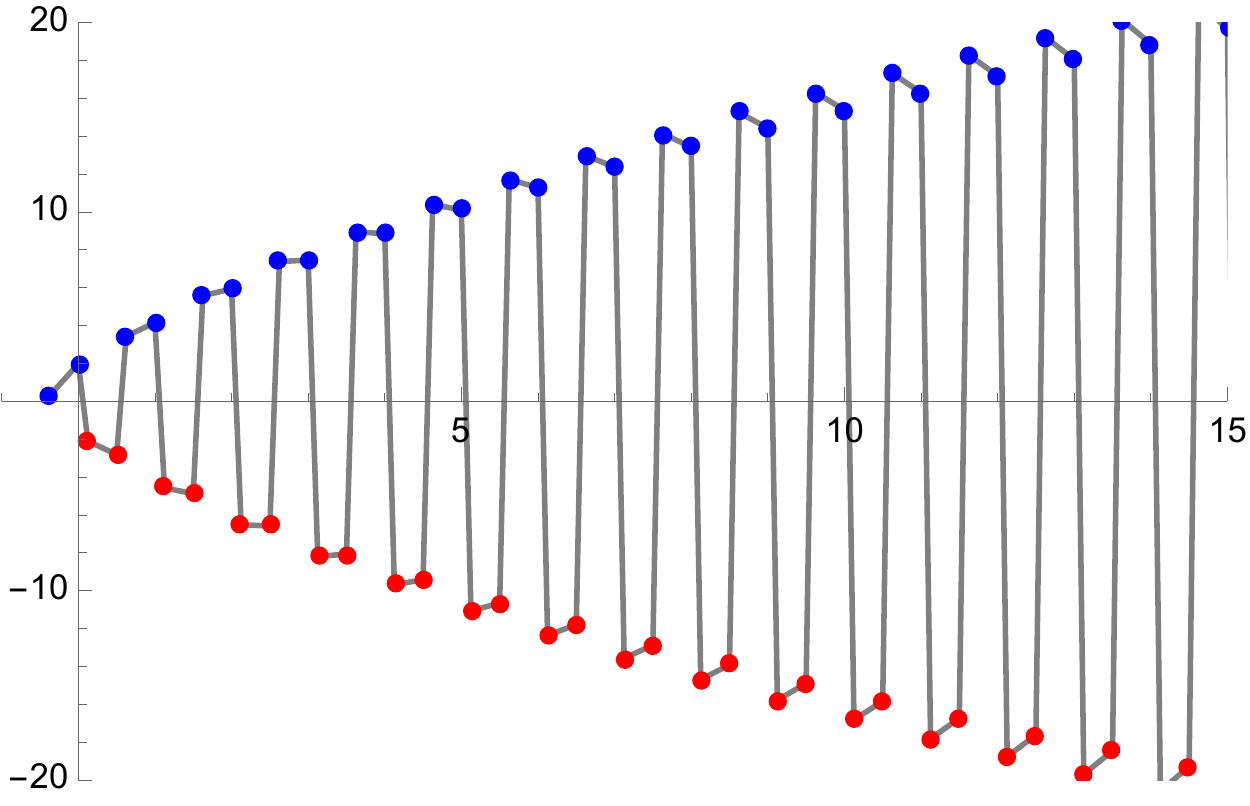"}
	\end{subfigure}$\qquad\qquad$
	\begin{subfigure}[b]{0.4\linewidth}	
	\includegraphics[width=\linewidth]{"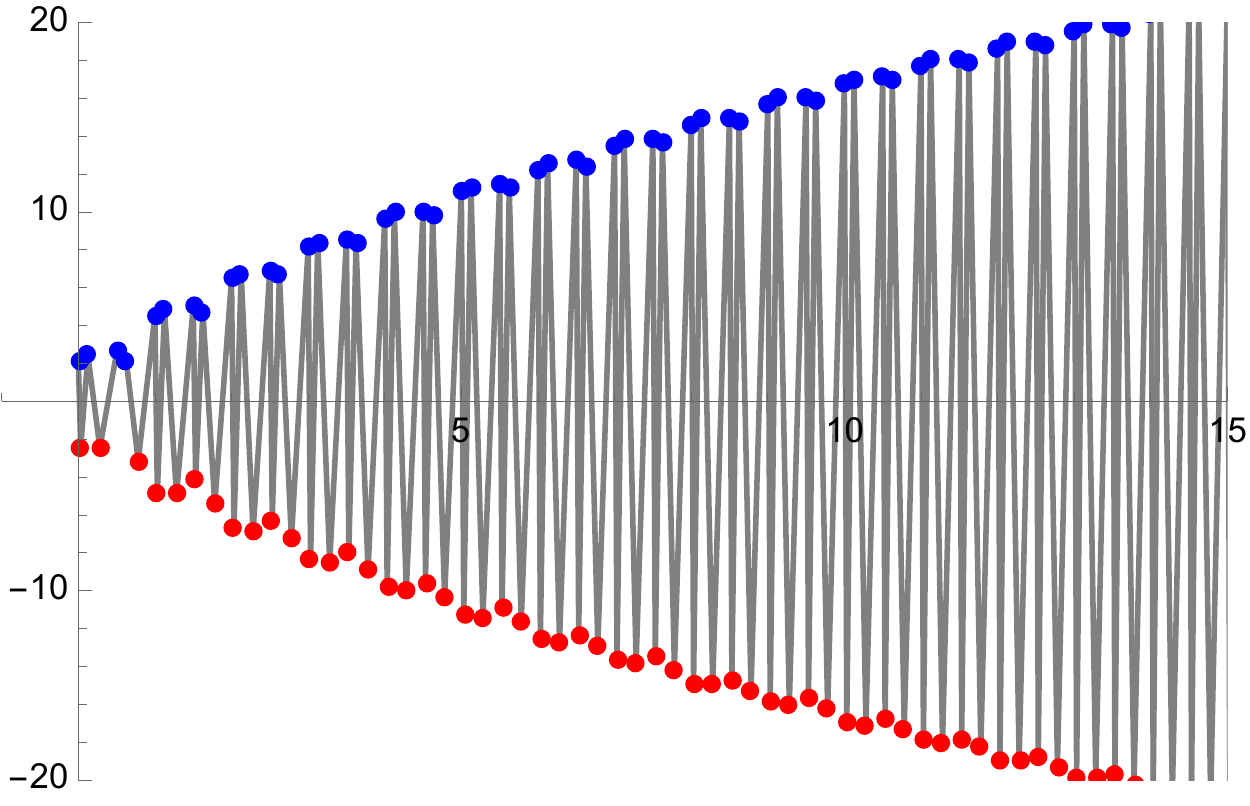"}
	\end{subfigure}
	\caption{The {\em signed} logarithm of the net number of degrees of freedom at each mass level for the type IIB Scherk-Schwarz reduction. Positive (negative) contributions are ascribed to the excess of bosonic (fermionic) states. The left figure refers to $R^2=2\, \alpha'$, and despite  the presence of tachyons,  the spectrum exhibits misaligned supersymmetry. The right figure corresponds to $R^2=8\, \alpha'$, within the non-tachyonic region. The step-like shape of the enveloping functions reflects the fact that the first few Kaluza-Klein excitations have masses smaller than the string scale, as first observed in \cite{Abel:2015oxa}.} \label{ScScplots}
\end{figure}

The effective degrees of freedom at a given mass level can be straightforwardly extracted from the partition function \eqref{ScScPart} by Taylor expanding the various characters. We find that misaligned supersymmetry is present for {\em any} choice of the compactification radius, even in the tachyonic regime, as shown in Figure \ref{ScScplots}. This is in accordance with our general discussion of Section \ref{main} and is corroborated by the large-$n$ behaviour of the sector averaged sum $\langle d(n)\rangle$. 

Indeed, the degeneracy of each character $\chi_a$ has in principle many exponential growth rates associated to all tachyonic characters of the RCFT. However, the leading exponential, associated to $\chi_0$, is universal and cancels in the sector averaged sum since the partition function involves $4s$ bosonic and fermionic sectors. Therefore, we can conclude that, for any $s$, $C_\text{eff} < C_\text{tot}$. Moreover, for $s>8$ no physical tachyons are present in the spectrum and thus $C_\text{eff} =0$. Finally, in the tachyonic region $s<8$, the state which dictates the exponential growth of $\langle d(n) \rangle$ is the physical tachyon $\chi_{s/2}\, \bar \chi_{3s/2}$ and its conjugate, so that
\begin{equation}
C_\text{eff} = 2 \pi \sqrt{ 8-s} \,. \label{CeffSS}
\end{equation}
Indeed, explicit calculations yield
\begin{equation}
\langle d(n) \rangle = \frac{81}{4096\, n^5} \left[ e^{2 \pi \sqrt{6n}} + e^{\pi \sqrt{6n}} + 2\, e^{2 \pi \sqrt{2n/3}} +\dots \right]\,,
\end{equation}
for $s=2$, 
\begin{equation}
\langle d(n) \rangle = \frac{1}{256\, n^5} \left[ e^{4 \pi \sqrt{n}} + e^{2 \pi \sqrt{n}} + 2\, e^{\frac{4}{3} \pi \sqrt{n}} +\dots \right]\,,
\end{equation}
for $s=4$, while $\langle d(n) \rangle = 0$  for $s>8$. 

\begin{figure}[h!]
	\centering
	\includegraphics[width=7cm]{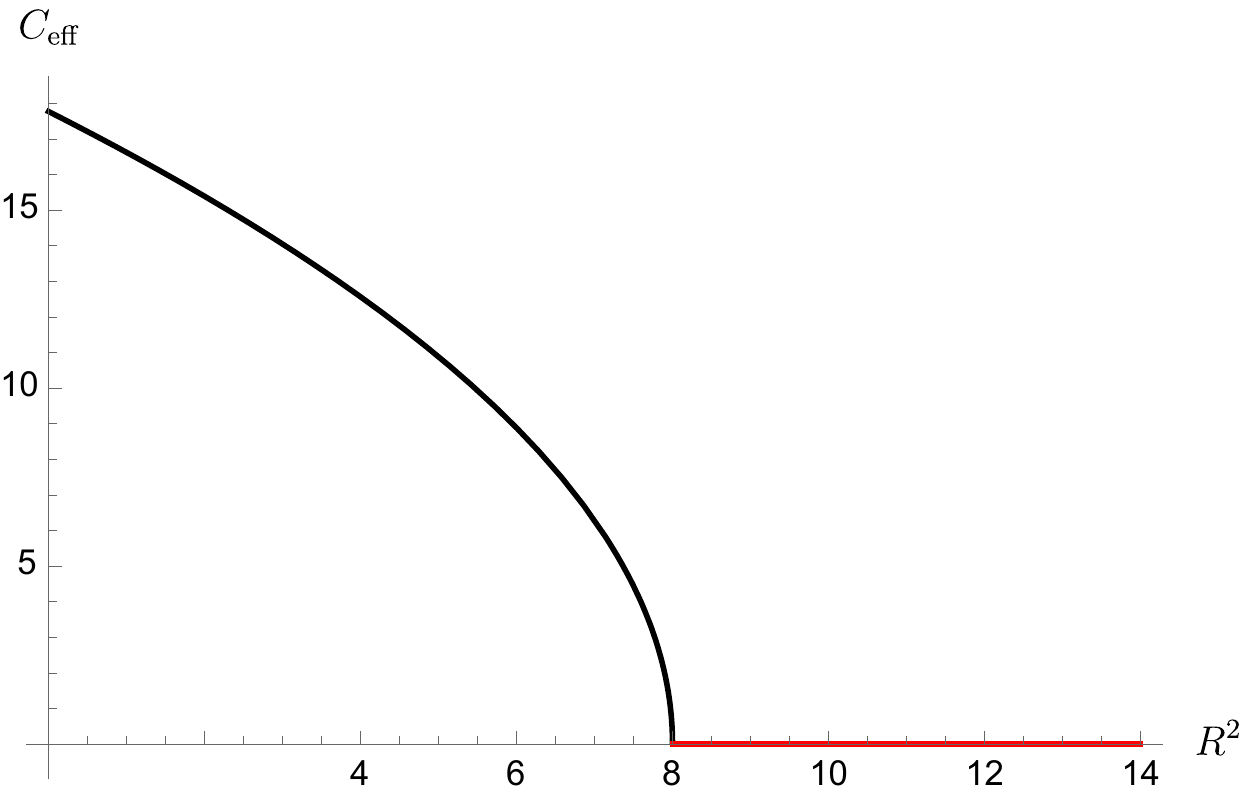}
	\caption{The  figure displays the dependence of the effective central charge on $R$ and shows that below the critical radius the Scherk-Schwarz reduction is not a deformation of the original theory.}\label{figCeff}
\end{figure}

It is tempting to continue the behaviour \eqref{CeffSS} to arbitrary irrational values of the compactification radius, so that the asymptotic growth of the {\em irrational} sector averaged sum reads
\begin{equation}
C_\text{eff} = \begin{cases}
2 \pi \sqrt{8-R^2/\alpha'} & \text{for}\quad R^2 <8\alpha '\,,
\\
 0 & \text{for}\quad  R^2 \ge 8\alpha' \,,
\end{cases} \label{CeffRirrat}
\end{equation}
as shown in Figure \ref{figCeff}, and $C_\text{eff} \to C_\text{tot}$ as $R\to 0$, in accordance with the fact that, in this limit, one recovers the purely bosonic type 0B theory.

\subsection{A Comment on Phase Transitions}

It is tempting to consider the partition function $\mathscr Z$ of the world-sheet CFT as a function of $q=e^{-\beta+i\mu}$, where we now interpret $\beta$ as the world-sheet inverse temperature and $\mu$ as the spin potential, conjugate to the worldsheet momentum operator. Clearly, the sum over states in $\mathscr Z$ is absolutely convergent, and the corresponding world-sheet free energy $\mathscr F = -(1/\beta) \log\mathscr Z$ is analytic, signifying the absence of phase transitions, as in any theory with a finite number of local degrees of freedom. For instance, in bosonic string theory the density of states grows universally according to the Cardy formula
\begin{equation}
	d(n) \sim e^{4\pi\sqrt{cn/24}} \,,
\end{equation}
so that the level matched partition function
\begin{equation}
	\int_0^1 d\tau_1 \,\mathscr Z(\tau_1, \beta/2\pi) = \sum_{n} d(n)\, e^{-2\beta n} \,,
\end{equation}
converges for all values of $\beta$. 

In light of our discussion on $C_\text{eff}$, it is interesting to construct a deformed version of the above partition function, obtained by analytically continuing the integer $n$ to the reals and by averaging the net degrees of freedom in terms of the enveloping functions $\Phi(n)$. This amounts to replacing the physical net degeneracies $d(n)$ by their sector averaged versions $\langle d(n)\rangle$, and defining a \emph{sector averaged partition function} for the world-sheet CFT as
\begin{equation}
	\langle \mathscr Z\rangle = \sum_n \langle d(n)\rangle \,e^{-2\beta n} \,.
\end{equation}
This deformation effectively introduces an infinite number of degrees of freedom by averaging the interpolation of the mass levels in terms of the enveloping functions $\Phi$. In doing so, one can estimate 
\begin{equation}
	\langle \mathscr Z\rangle \sim \int dn \, e^{C_\text{eff}\sqrt{n} -2\beta n} \sim e^{\frac{C_\text{eff}^2}{8\beta}}\,,
\end{equation}
so that the \emph{sector averaged free energy} reads
\begin{equation}
	\langle \mathscr F\rangle \sim - \frac{C_\text{eff}^2}{8\beta^2} \,,
\end{equation}
and is controlled by the square of the effective central charge. In the case of Scherk-Schwarz supersymmetry breaking, $C_\text{eff}$ is a function of the compactification radius $R$, as shown in eq. \eqref{CeffRirrat}, which, from the point of view of the worldsheet CFT, should be treated as an external background field. In this sense, the sector averaged free energy $\langle \mathscr F\rangle$ displays a first order phase transition as the radius crosses the critical value $R_c$, since its first derivative is discontinuous.

This would suggest a possible interpretation in terms of phase transitions in a suitable holographic dual. In fact, it is known that two-dimensional CFT's may admit an $\text{AdS}_3$ gravitational description, and in certain deformations of symmetric orbifold CFTs, in the large central charge limit,  Hagedorn-like phase transitions of the CFT are mapped to Hawking-Page transitions of the gravity theory, dominated by the entropy of BTZ black holes \cite{Keller:2011xi}. It is thus tempting to interpret the phase transition displayed by our system in terms of a holographic dual where a similar kind of averaging procedure is introduced. However, this investigation lies beyond the scope of this work.

\section{Conclusions}\label{concl}

In this paper we have revisited the problem of classical stability, {\em i.e.} absence of on-shell tachyons, of non-supersymmetric string vacua in various dimensions in relation to the growth rate of the net number of degrees of freedom. We find that boson/fermion oscillations is not an exclusive property of non-tachyonic strings, contrary to certain misconceptions in the literature, but manifests itself in all theories, tachyonic or not, containing space-time fermions. In fact, we have shown that the growth rate of the sector averaged sum $\langle d (n) \rangle$ is set by the mass of the lightest state, whether it be the deepest tachyon or massless particles, and is strictly  smaller than $C_\text{tot}$ in the presence of fermions. We have also proven that the necessary and sufficient condition for the tree-level stability of the vacuum is the vanishing of  $\langle d (n) \rangle$ corresponding to $C_\text{eff} =0$, as conjectured in \cite{Dienes:1994np}. Our result is model independent  and applies to any closed-string vacuum in any dimension, and agrees with the recent analysis of \cite{Cribiori:2020sct} conducted for the $\text{SO} (16) \times \text{SO} (16)$ heterotic string. 

Following \cite{Dienes:1994np,Cribiori:2020sct}, our analysis is based on the sector average of the string degrees of freedom whose discrete masses are analytically continued to real values. The cancellations required by misaligned supersymmetry highly depend on this analytic continuation and not just on the properties of the discrete spectrum.
This is to be contrasted to the works of \cite{Kutasov:1990sv} and \cite{Angelantonj:2010ic} which, using number theoretic methods, relate misaligned supersymmetry to the real physical degrees of freedom of the vacuum, without having to resort to an analytic continuation of the mass levels. It would be interesting to obtain a more direct connection between these two {\em a priori} different approaches. 

Any quantitative analysis on misaligned supersymmetry heavily relies on modular invariance of the torus partition function, and can thus only be formulated for closed oriented strings. Still orientifold vacua \cite{Sagnotti:1987tw, Horava:1989vt, Pradisi:1988xd, Bianchi:1991eu, Bianchi:1990yu} provide an appealing phenomenology and afford more general ways to break supersymmetry \cite{Sagnotti:1995ga, Sagnotti:1996qj, Angelantonj:1998gj, Antoniadis:1998ki, Sugimoto:1999tx, Antoniadis:1999xk, Aldazabal:1999jr, Angelantonj:1999jh, Angelantonj:1999ms, Angelantonj:2000xf,Dudas:2004vi} which cannot be realised in closed strings. Although some attempts have been made to understand the role played by misaligned supersymmetry in the classical stability of non-supersymmetric orientifolds \cite{Niarchos:2000kw, Israel:2007nj, Cribiori:2020sct, Cribiori:2021txm}, a thorough quantitative analysis is still lacking. Modular transformations relate in them the direct and transverse channels, so that there is no obvious link between the IR and the UV properties of the spectrum, and therefore different tools need to be employed to uncover this connection. We hope to return to this problem in the near future.

\section*{Acknowledgements}
It is a pleasure to thank Ivano Basile, Flavio Tonioni and in particular Niccol\`o Cribiori for enlightening discussions on misaligned supersymmetry and non-supersymmetric string vacua. We are grateful to Augusto Sagnotti for constructive feedback on the manuscript and especially to Keith Dienes for precious comments on the first version of the manuscript and for clarifications on the essence of misaligned supersymmetry. I.F. would like to thank the Physics Department of the University of Torino for hospitality during the final stages of this project. C.A. and G.L. would like the Department of Theoretical Physics at CERN for hospitality during various stages of this project. The work of C.A. is partially supported by the MIUR-PRIN contract 2017CC72MK-003.

\begin{appendix}

\section{Rademacher Expansion for Vector-Valued Modular Forms}\label{appRadexp}

In this Appendix we derive an exact expression for the Fourier coefficients of vector-valued modular forms of weight $w \le 0$ with a multiplier system, which first appeared in \cite{Dijkgraaf:2000fq, Manschot:2007ha}. Our presentation extends the discussion of \cite{Sussman2017} valid for $\eta$-quotients. Let us consider vector-valued modular forms $f_a (\tau )$, $a=1, \dots, M $, transforming under the action of the modular group as
\begin{equation}
f_a (\gamma\cdot \tau ) = (c\tau + d )^w\, M(\gamma )_{ab} \, f_b (\tau )\,, \label{vvmftr}
\end{equation}
where $\gamma$ is a generic element of $\text{SL} (2;\mathbb{Z})$, $w$ is the weight of the modular forms that we take to be negative and not necessarily integer, $M(\gamma )$ is a unitary matrix, including a non-trivial multiplier system, representing the action of $\gamma$ on the $f$'s, and a sum over the repeated index $b$ is understood. These vector-valued modular forms play a crucial role in string theory, since they represent the pseudo characters of weight $w=1-D/2$ we have introduced in Section \ref{KV},
\begin{equation}
f_a (\tau ) = \hat\chi_a (\tau) = \eta^{2-D}\, \chi_a (\tau )\,.
\end{equation}
In this case, the multiplier system is associated to the Dedekind $\eta$ functions and, for the transformation
\begin{equation}
\gamma = \begin{pmatrix} a & b \\ c & d \end{pmatrix} \in \text{SL} (2;\mathbb{Z} )\,,
\end{equation}
reads
\begin{equation}
\zeta (\gamma ) = \begin{cases}
e^{\frac{i \pi b}{6} w} & \text{for}\quad c=0,\ d=1\,,
\\
e^{-\frac{i \pi b}{6} w} & \text{for}\quad c=0,\ d=-1\,,
\\
e^{2 \pi i w \left( \frac{a+d}{12 c}- s(d,c) - \frac{1}{4} \right)} & \text{for}\quad c>0\,,
\\
e^{2 \pi i w \left( \frac{a+d}{12 c}- s(-d,-c) - \frac{1}{4} \right)} & \text{for}\quad c<0\,,
\end{cases}
\end{equation}
with 
\begin{equation}
s(c,d) = \sum_{n=1}^{c-1} \frac{n}{c} \left( \frac{nd}{c} - \left\lfloor\frac{nd}{c} \right\rfloor- \frac{1}{2} \right)\,,
\end{equation}
the Dedekind sum. The action of $M (\gamma)$ on the space of characters is then given by suitable combinations of the $S$ and $T$ matrices of the CFT. The phases induced by the multiplier system are precisely the phases that enter in the definition of the hatted matrices in \eqref{STmulti}.

The vector-valued modular forms (or the pseudo-characters for what matters) admit a Fourier series expansion 
\begin{equation}
f_a (\tau ) = \sum_{n=0}^\infty d_a (n) \, q^{n+H_a}\,,
\end{equation}
with $q=e^{2\pi i \tau}$. The Fourier coefficients are formally given by
\begin{equation}
d_a (n) = \oint_\Gamma \frac{dq}{2\pi i} \frac{f_a (q)}{q^{n +1 +H_a}}\,,
\end{equation}
with $\Gamma$ a closed contour encircling the origin $q=0$ and entirely contained within the unit disk $|q|=1$, which is the domain of definition of the functions $f_a (q)$. This integral can  be computed by summing the contributions of the infinite essential singularities which lie on the boundary of the unit disk, $|q|=1$. In the $\tau$ plane these are associated to the rational points along the real axis, within a strip of width one, and are the images of the cusp $i\infty$ under the action of the modular group. Hardy and Ramanujan \cite{Hardy} and Rademacher \cite{Rademacher:1937a, Rademacher:1937b, Rademacher:1938} have devised a powerful method, known as the {\em circle method}, to compute the integral which amounts in a suitable choice of the contour $\Gamma$. To this end, it is convenient to introduce the Farey fractions ${\mathscr F}_N$ of order $N$. They are sequences of rational numbers $p/\ell$ with $(p,\ell)=1$, $p,\ell \le N$, and $0\le p/\ell <1$ arranged in increasing order. Given a fraction $p/\ell$, its immediate predecessor or successor are the fractions $p_p/\ell_p$ and $p_s/\ell_s$ which precede and succeed $p/\ell$ in ${\mathscr F}_N$. Two consecutive elements $p/\ell <p'/\ell'$ in a Farey fraction satisfy $\ell p'-\ell 'p=1$. Notice that ${\mathscr F}_N \supset {\mathscr F}_{N-1}$ and therefore in the limit $N\to \infty$ the Farey sequence contains all the essential singularities of $f_a (\tau )$. 

To a given pair $(p,\ell)$ of co-prime integers with $0\le p < \ell$, we can associate a Ford circle $C (p,\ell)$ of radius $1/2\ell^2$ centred at the point
\begin{equation}
\frac{p}{\ell} + \frac{i}{2 \ell^2}\,.
\end{equation}
If $p/\ell \in {\mathscr F}_N$, we define the upper arc $\gamma (p,\ell)$ of the Ford circle to be the portion of $C(p,\ell)$ from the initial point
\begin{equation}
\tau_I (p,\ell) = \frac{p}{\ell} - \frac{\ell_p}{\ell (\ell^2 + \ell_p^2)}+ \frac{i}{\ell^2 + \ell_p^2}\,,
\end{equation}
to the terminal point
\begin{equation}
\tau_T (p,\ell) = \frac{p}{\ell} + \frac{\ell_s}{\ell (\ell^2 + \ell_s^2)}+ \frac{i}{\ell^2 + \ell_s^2}\,,
\end{equation}
traversed in the clockwise direction. The Ford circles associated to two elements of a Farey sequence are either tangent if the associated fractions are adjacent in ${\mathscr F}_N$ or do not intersect. 

The path $p(N)$, of order $N$, used in the circle method is then the union of all arcs $\gamma (p,\ell)$ associated to all fractions in the Farey sequence ${\mathscr F}_N$,
\begin{equation}
p (N) = \bigcup_{p/\ell\in {\mathscr F}_N} \gamma (p,\ell)\,.
\end{equation}
Notice that, on the unit strip, this is a closed path since $\tau_I (0,1) = \tau_T (N-1,N)$, if we view $0/1$ as the immediate successor of $(N-1)/N$.

The Fourier coefficients of $f_a$ can then be written as
\begin{equation}
\begin{split}
d_a (n) &= \lim_{N\to \infty} \oint_{p(N)} d\tau\, f_a (\tau ) \, e^{-2\pi i \tau (n+ H_a)} 
\\
&= \lim_{N\to \infty} \sum_{p/\ell \in {\mathscr F}_N} \int_{\gamma (p,\ell)} d\tau\, f_a (\tau ) \, e^{-2\pi i \tau (n+ H_a)} 
\\
&= \lim_{N\to \infty} \sum_{\ell=1}^N \sum_{p=0\atop (p,\ell)=1}^{\ell-1}  \int_{\gamma (p,\ell)} d\tau\, f_a (\tau ) \, e^{-2\pi i \tau (n+ H_a)}  \,.
\end{split}
\end{equation}
It is convenient to perform the change of variable
\begin{equation}
\tau = \frac{p}{\ell} + i \frac{z}{\ell^2}\,,
\end{equation}
which maps each Ford circle to the unique circle $S_{1/2} \left(\frac{1}{2} \right)$ in the $z$-plane, centred at $z=\frac{1}{2}$ with radius $1/2$. Each arc $\gamma (p,\ell)$ is then mapped to the arc $\zeta (z_I,z_T)$ on $S_{1/2} \left(\frac{1}{2} \right)$ starting at $z_I$ and terminating at $z_T$, with
\begin{equation}
z_I = \frac{\ell^2 + i \ell_p \, \ell}{\ell^2 + \ell_p^2} \,,
\qquad
z_T =\frac{\ell^2 - i \ell_s\, \ell}{\ell^2 + \ell_s^2} \,.
\end{equation}
As a result,
\begin{equation}
d_a (n) = \lim_{N\to \infty}  \sum_{\ell=1}^N \sum_{p=0\atop (p,\ell)=1}^{\ell-1} \frac{i\, e^{-2\pi i  \frac{p}{\ell}(n+ H_a)}}{\ell^2}\,  \int_{\zeta (z_I,z_T)} dz\, f_a \left(\frac{p}{\ell} + i \frac{z}{\ell^2} \right) \, e^{\frac{2\pi}{\ell^2} z (n+ H_a)} \,.
\end{equation}
Since the new contour does not cross the point $z=0$, which corresponds to the position of the essential singularities $\tau = p/\ell$, one can freely deform it and replace the arcs $\zeta (p,\ell)$ by the chords $z (p,\ell)$ connecting the end points of the arcs and contained entirely inside the disk $S_{1/2} \left( \frac{1}{2}\right)$. Points on the chord satisfy $|z| \le \sqrt{2} \ell / N$ so that the length of the chord is at most $2\sqrt{2} \ell /N$ \cite{Sussman2017}. One can see now the power of the circle method, since in the limit $N\to \infty$ where the contributions of all singularities are taken into account, the point $z\to 0$, and one can use the modular properties of the vector-valued modular forms to connect their behaviour near the rational points in the $\tau$ plane to their value at the cusp $\tau = i \infty$. This map corresponds to the modular transformation
\begin{equation}
\gamma_{p,\ell} = \begin{pmatrix} - p' & \frac{1 + pp'}{\ell} \\ -\ell & p\end{pmatrix} \,, \qquad \gamma_{p,\ell}: \frac{p}{\ell} + i\frac{z}{\ell^2} \to \frac{p'}{\ell} + \frac{i}{z}\,,
\end{equation}
with $p'$ fixed by the condition that $\gamma_{p,\ell}\in \text{SL} (2;\mathbb{Z})$. Therefore, using \eqref{vvmftr}, 
\begin{equation}
d_a (n) = \lim_{N\to \infty} i^{1+w} \sum_{\ell=1}^N \sum_{p=0 \atop (p,\ell )=1}^{\ell-1} \ell^{w-2}\, e^{-2\pi i  \frac{p}{\ell}(n+ H_a)}\,  (M^{-1}_{p,\ell} ) _{ab} \, \int_{z (p,\ell)} dz\, z^{-w}\, f_b \left( \frac{p'}{\ell} + \frac{i}{z} \right) \, e^{\frac{2\pi}{\ell^2}\, z \, (n+H_a)} \,.
\end{equation}
As $N$ increases, the dominant contribution comes from the {\em lightest states} with $m=0$ in the Fourier expansion of $f_b$, so that we can formally write
\begin{equation}
d_a (n) = \lim_{N\to \infty} \left( I_N + R_N + J_N \right)\,,
\end{equation}
where 
\begin{equation}
I_N = i^{1+w} \sum_{\ell =1}^N \, \sum_{p=0\atop (p,\ell )=1}^{\ell -1} \ell^{w-2}\, (M^{-1}_{\ell ,p})_{ab} d_b (0)\, e^{\frac{2\pi i}{\ell} \left( H_b p' - p (n+H_a )\right)}\, \int_{S_{1/2} \left(\frac{1}{2}\right)} dz\, z^{-w} \, e^{-2\pi \left( \frac{H_b}{z} - \frac{z (n+H_a)}{\ell^2}\right)}\,,
\end{equation}
is the dominant term, and $R_N$ and $J_N$ are the error terms. Notice that in writing $I_N$ we have also deformed the contour of integration, that is why we have two different types of error. The first
\begin{equation}
R_N = i^{1+w} \sum_{\ell =1}^N \sum_{p=0\atop (p,\ell )=1}^{p-1} \sum_{m=1}^\infty \, \ell^{w-2} (M^{-1}_{\ell ,p})_{ab} d_b (m) \, e^{\frac{2\pi i}{\ell} \left(  (m+H_b) p' -  (n+H_a ) p \right)}\, \int_{z (p , \ell)}dz\, z^{-w} \, e^{-2\pi \left( \frac{m+H_b}{z}- \frac{z(n+H_a )}{\ell^2} \right)}
\end{equation}
takes into account the fact that we include in $I_N$ only the zero mode of the Fourier expansion of the modular form. The second
\begin{equation}
J_N = i^{1+w} \sum_{\ell=1}^N \sum_{p=0\atop (p,\ell )=1}^{\ell -1} \ell^{w-2}\, (M^{-1}_{\ell ,p})_{ab} \, d_b (0) \, e^{\frac{2\pi i}{\ell} (H_b \, p' -(n+H_a )p )}
\int_{\zeta (0, z_I ) \cup \zeta (z_T , 0)} dz\, z^{-w} \, e^{-2\pi \left( \frac{H_b}{z}-\frac{z(n+H_a)}{\ell^2} \right)}
\end{equation}
takes into account the fact that we have deformed the integration contour, since $\zeta (0, z_I ) \cup \zeta (z_T , 0)\cup z (p,\ell) = S_{1/2} \left( \frac{1}{2}\right)$.

Let us concentrate on the dominant term first. The integral can be computed by {\em opening} the circe $S_{1/2} \left( \frac{1}{2} \right)$ into the vertical line $(1-i\infty , 1+i \infty)$ via the change of variable $y = \frac{\ell}{z} \sqrt{\frac{|H_b|}{n+H_a}}$. Jordan's lemma implies that it is non-vanishing only when $H_b <0$, and in this case is proportional to the modified Bessel function of the first kind $I_\nu (z)$. Altogether,
\begin{equation}
I_N = 2\pi \, \sum_{\ell =1}^N  \sum_{b\, | \, H_b <0}\, Q^{(\ell , n)}_{ab} d_b (0) \, \ell^{-1} \left( \frac{|H_b|}{n+H_a}\right)^{\frac{1-w}{2}}\, I_{1-w} \left( \frac{4\pi}{\ell} \sqrt{ |H_b| (n+H_a )}\right)\,,
\end{equation}
where now the sum over $b$ is restricted to those $f$'s that have negative $H_b$, and 
\begin{equation}
Q ^{(\ell , n)}_{ab} = i^w \, \sum_{p=0\atop (p,\ell )=1}^{\ell -1} e^{\frac{2\pi i}{\ell} \left( H_b p' - p (n+H_a )\right)}\, (M^{-1}_{\ell ,p})_{ab}
\end{equation}
is a generalisation of the Kloosterman sum. 

We now move to estimate of the error terms. Starting from $R_N$, we focus on what is left from the Fourier series,
\begin{equation}
\Delta_b^{\ell, p} \equiv \sum_{m=1}^\infty d_b (m) e^{2\pi i (m+H_b) \left( \frac{p'}{\ell} + \frac{i}{z}\right)}\,.
\end{equation}
We have
\begin{equation}
|\Delta_b^{\ell , p}| \le d_b (1)\, e^{-2\pi (1+H_b ) \text{Re} (1/z)} + \sum_{m=2}^\infty d_b (m) e^{-2\pi (m+ H_b ) \text{Re} (1/z)} \le D \, e^{-2\pi (1+H_b ) \text{Re} (1/z)}\,,
\end{equation}
where $D$ is a finite constant. The bound arises because, following Hardy and Ramanujan, the coefficients $d_b (m) \sim O( e^{\pi \sqrt{2 c m /3}} )$ and therefore the series is absolutely and uniformly convergent. Moreover, $\text{Re} (1/z ) \ge 1$ since all the chords $z (p, \ell)$ are inside the circle $S_{1/2} \left( \frac{1}{2}\right) $, so that
\begin{equation}
|R_N | \le 2\, \sqrt{M} \, D\, e^{-2\pi (1-n+H_b-H_a)}\, \sum_{\ell=1}^N \sum_{p=0\atop (p,\ell )=1}^{\ell -1} \ell^{w-2} \, \left( \frac{\sqrt{2} \ell}{N}\right)^{1-w}\,.
\end{equation}
Here we have also used the bound $\sum_b |M_{p,\ell} | \le \sqrt{M}$, with $M$ the total number of modular forms $f_b$ \cite{Kani:1989im}, and the fact that on the chord $|z|\le \sqrt{2}\ell/N$. Finally, the sum over $p$ yields the Euler totient function $\varphi (\ell )$ which is bounded by $\ell$, so that
\begin{equation}
|R_N| \le 2^{\frac{3-w}{2}}    \, \sqrt{ M} \, D\, e^{2\pi (n-1+H_a-H_b)}\, N^w \underset{N\to \infty}{\longrightarrow}  0
\end{equation}
since $w<0$. 

As for the error $J_N$ we split $J_N = J_N^I + J_N^T$ each contribution associated to the two arcs in the integration domain. Following similar steps, and taking into account that on the arc $|z|\le \pi \sqrt{2} \ell/N$, one finds
\begin{equation}
|J^{I,T}_N| \le 2^{\frac{1-w}{2}}\, \pi \sqrt{M}\, d_b (0)\, e^{2\pi (n+H_a - H_b )}\, N^w \underset{N\to \infty}{\longrightarrow}  0 \,.
\end{equation}

Taking this into account, we conclude that the {\em exact} expression for the Fourier coefficients of vector-valued modular forms of negative weight $w$ is
\begin{equation}
d_a (n) = 2\pi \, \sum_{\ell=1}^\infty \sum_{b\, | \, H_b <0}\, Q^{(\ell , n)}_{ab} \, d_b (0) \, \ell^{-1} \, \left( \frac{|H_b|}{n+H_a}\right)^{\frac{1-w}{2}}\, I_{1-w} \left( \frac{4\pi}{\ell} \sqrt{ |H_b| (n+H_a )}\right) \,.\label{finalRes}
\end{equation}
In this expression we have assumed that $|H_b|\le 1$. Although this is the case of interest in string theory, one can easily modify the previous expression whenever $|H_b| \le h$, with $h$ a positive integer. In this case, the sum over $b$ includes all Fourier modes $d_b (m)$ which have $m+H_b <0$ and $|H_b|\to |m+H_b|$. 

For zero weight one needs a more precise treatment. In the case of modular functions, using a bound of Davenport on the Kloosterman sum, Rademacher showed that the error is of order $N^{-1/3}$, thus extending the range of validity of the procedure to the case $w=0$ \cite{Rademacher:1938b}. A similar refined estimate is believed to exist \cite{Dijkgraaf:2000fq, Manschot:2007ha} also for the generalised Kloosterman sums which would make eq. \eqref{finalRes} true also for the case $w=0$, of interest for two-dimensional string vacua.

\section{The Scherk-Schwarz Mechanism in RCFT's }\label{appScScrational}
	
Although the Narain partition function for a real boson $Y$ compactified on a circle $S^1 (R)$ of radius $R$ does not fully factorise into the product of holomorphic and anti-holomorphic contributions, things simplify considerably whenever $R^2/\alpha' = s/t\in \mathbb{Q}$ takes rational values. In this case, the CFT of the compact boson becomes rational, the Kaluza-Klein momenta and windings admit the parametrisation
\begin{equation} \label{momenta}
 m= s(k + \bar{k}) + \tfrac{1+l}{2t} \alpha \,, \qquad  n= t(k - \bar{k}) + \tfrac{1-l}{2s} \alpha \,,
\end{equation}
with $l$ defined after eq. \eqref{defl} and $k,\bar k \in \mathbb{Z}$, and only a finite number $N=2 st$ of representations are unitary, and are associated to the characters
\begin{equation}\label{chrla}
	\lambda_{\alpha}= \frac{1}{\eta}\, \sum_{k \in \mathbb{Z}} q^{\frac{N}{2}( k + \frac{\alpha}{N})^2}.
\end{equation}
Although the $\lambda$'s provide a natural decomposition of the Narain partition function \eqref{NarainL} as the sesquilinear combination \eqref{NarainRCFT}, in general they do not provide a suitable basis when shift orbifolds act on $S^1 (R)$. In fact, already for the simple order-two shift  $\delta :\  Y\to Y+\pi R$, the Narain lattice picks up a phase
\begin{equation}
\sum_{m,n} \Lambda_{m,n} \to \sum_{m,n} (-1)^m \, \Lambda_{m,n}
\end{equation}
and in view of \eqref{momenta}, it is clear that the characters $\lambda_\alpha$ are eigenstates of $\delta$ only for even $s$ so that $(-1)^m \to (-1)^{\tfrac{1+l}{2t} \alpha}$. For odd $s$, instead, the $k$-th excitation in \eqref{chrla} acquires an additional sign depending on the parity of $k$, that would suggest the decomposition
\begin{equation}
\lambda_\alpha  \to \xi_\alpha^i = \frac{1}{\eta} \sum_{k \in \mathbb{Z}} q^{\frac{N}{2}\left( 2 k + i + \frac{\alpha}{N}\right)^2}\,, \qquad i=0,1\,.
\end{equation}
The $\xi$'s however, are not closed under the action of $\text{SL} (2,\mathbb{Z})$, since they fail to capture the {\em twisted} sector with half-integer windings. Therefore, the correct choice of the $\delta$ eigenstates is
\begin{equation}
\zeta_\alpha=  \frac{1}{\eta }\, \sum_{k} q^{2 N \left ( k + \frac{\alpha}{4 N} \right )^2} 
\end{equation}
with now $\alpha =0, \ldots , 4 N-1$. 

One can thus decompose the various orbifold blocks as 
\begin{equation}
\begin{split}
\sum_{m,n} \Lambda_{m,n} &=  \begin{cases} \sum_{\alpha =0}^{N-1} \lambda_\alpha \, \bar \lambda_{l \alpha } & \text{for even $s$} \,,
\\
\sum_{a,b=0}^1 \sum_{\alpha=0}^{N-1} \zeta_{2(\alpha + a N)} \bar{\zeta}_{2(l \alpha + b N)} & \text{for odd $s$}\,,
\end{cases}
\\
\sum_{m,n} (-1)^m \, \Lambda_{m,n} &= \begin{cases} \sum_{\alpha =0}^{N-1} (-1)^{\frac{1+l}{2t} \alpha}\, \lambda_\alpha \, \bar \lambda_{l \alpha } & \text{for even $s$} \,,
\\
\sum_{a,b=0}^1 \sum_{\alpha=0}^{N-1} (-1)^{a+b + \frac{1+l}{2t} \alpha} \, \zeta_{2(\alpha + a N)} \bar{\zeta}_{2(l \alpha + b N)} & \text{for odd $s$} \,,
\end{cases}
\\
\sum_{m,n} \Lambda_{m,n+\frac{1}{2}} &= \begin{cases} \sum_{ \alpha=0}^{N-1}  \lambda_{l \alpha -(1+l)s/2} \bar{\lambda}_{\alpha}
& \text{for even $s$} \,,
\\
\sum_{c=0}^1 \sum_{\alpha=0}^{2N-1}  \zeta_{l(2\alpha+1)-(1+l)s + 2 N c} \ \bar{\zeta}_{2 \alpha + 1} & \text{for odd $s$}\,,
\end{cases}
\\
\sum_{m,n} (-1)^m \, \Lambda_{m,n+\frac{1}{2}} &= \begin{cases} \sum_{\alpha=0}^{N-1} (-1)^{\frac{(l \alpha -\frac{1+l}{2}s)^2 - \alpha^2}{N}} \lambda_{l \alpha -(1+l) s /2} \bar{\lambda}_{\alpha}
& \text{for even $s$} \,, \\
\sum_{c=0}^1 \sum_{\alpha=0}^{2N-1}  (-1)^{\frac{(l(2\alpha+1)-(1+l)s + 2 N c)^2 - (2 \alpha +1)^2}{4 N}}
			\zeta_{l(2\alpha+1)-(1+l)s + 2 N c} \ \bar{\zeta}_{2 \alpha + 1}  & \text{for odd $s$} \,.
\end{cases}
\end{split}
\end{equation}
Using the representation of the $\text{SL}(2,\mathbb{Z})$ generators 
\begin{equation}
T_{\alpha \beta}= e^{i \pi \left (\frac{\alpha^2}{M}-\frac{1}{12} \right )} \,, \qquad  S_{\alpha \beta}= \frac{e^{ \frac{2 \pi i \alpha \beta}{M}}}{\sqrt{M}} \,,
\end{equation}
on the space of characters, with $M=N$ for the $\lambda$'s,  and $M=4N$ for the $\zeta$'s, it is straightforward  to show that the relations
\begin{equation}
\sum_{m,n} (-1)^m \, \Lambda_{m,n}  \quad \overset{S}{\longrightarrow} \quad 
\sum_{m,n} \Lambda_{m,n+\frac{1}{2}} \quad \overset{T}{\longrightarrow} \quad  \sum_{m,n} (-1)^m \, \Lambda_{m,n+\frac{1}{2}} 
\end{equation}
hold both for odd and even $s$. 

\end{appendix}

\newpage

\bibliographystyle{unsrt}

\end{document}